\documentclass[11pt]{article}
\usepackage{geometry}
\usepackage{cite}
\usepackage{amsmath}
\usepackage{amssymb}
\usepackage{amsfonts}
\usepackage{graphicx}
\usepackage{caption}
\usepackage{subcaption}
\usepackage{dcolumn}
\usepackage{float}
\usepackage[usenames,dvipsnames]{xcolor}
\usepackage{tikz}
\usepackage{lipsum}
\usepackage{appendix}
\usepackage{authblk}
\usepackage{verbatim}
\usepackage{setspace}
\usepackage{gensymb}
\usepackage{tikz}
\usepackage{hyperref}
\usepackage{isotope}
\usepackage{booktabs}
\usepackage{tabularx}
\usepackage{caption}
\usepackage{adjustbox}
\usepackage{subcaption}
\hypersetup{
	colorlinks=true,
	linkcolor=blue,
	filecolor=cyan,      
	urlcolor=magenta,
}
\begin{document}

\title{\textbf{Tidal disruption of white dwarfs in a modified gravity theory with SPH} }
\date{}

\author[1]{Debojyoti Garain\thanks{dgarain@iitk.ac.in}}
\author[2]{Pritam Banerjee\thanks{pritam@phy.iitkgp.ac.in}}
\author[1]{Shaswata Chowdhury\thanks{shaswata@iitk.ac.in}}
\author[1]{Tapobrata Sarkar\thanks{tapo@iitk.ac.in}}

\affil[1]{Department of Physics, Indian Institute of Technology Kanpur, Kanpur 208016, India}
\affil[2]{Department of Physics, Indian Institute of Technology Kharagpur, Kharagpur 721302, India.}

\maketitle
\begin{abstract}
Low energy imprints of modifications to general relativity are often found in pressure balance equations inside stars. These modifications are then amenable to 
tests via astrophysical phenomena, using observational effects in stellar astrophysics that crucially depend on such equations. One such effect is tidal disruption of stars in the vicinity of black holes. In this paper, using a numerical scheme modelled with smoothed particle hydrodynamics, we study real time tidal disruption of a class of white dwarfs by intermediate-mass black holes, in the low energy limit of a theory of modified gravity that alters the internal physics of white dwarfs, 
namely the Eddington inspired Born-Infeld theory. In this single parameter extension of general relativity, 
the mass-radius relation of white dwarfs as well as their tidal disruption radius depend on the modified gravity parameter, and these capture the effect of modifications to general relativity. 
Our numerical simulations incorporating these show that departure from general relativity in these scenarios might be observationally significant, and should 
therefore be contrasted with data. In particular, we study observationally relevant
physical quantities, i.e., tidal kick velocity  and trajectory deviation of the remnant core and fallback rates of the tidal debris in this theory and compare them to the Newtonian limit of general relativity. We also comment on 
the qualitative differences between the modified gravity theory and one with stellar rotation. 

\end{abstract}
	
\section{Introduction}

Einstein's theory of general relativity (GR) remains the most successful theory of gravity till date. However, it is now widely believed that
modifications to GR are probably necessary and unavoidable. On the phenomenological side, issues of
cosmic acceleration and the cosmological constant have motivated many groups to search for possible modifications of GR, especially as an 
alternative explanation to dark energy, for recent reviews see, e.g.  \cite{CliftonRev, NOORev, LangloisRev, IshakRev, KaseRev}.  
At a more fundamental level lies the issue of singularities. Indeed, singularities are known to arise
in gravitational collapse processes, a paradigmatic example being that of a star with its core fuel exhausted, that collapses to a black hole. 
Singularities, often manifest as geodesic incompleteness of space-times (see e.g. \cite{Penrose, Hawking}), signal a breakdown or the limits of applicability of GR. 
(for comprehensive overviews, see \cite{HawkingEllis, Wald} or the more recent review in \cite{Senovilla}).
Singularities have been studied ever since the inception of GR, and in spite of several celebrated works, the issue of their resolution remains unclear. 
In this context, it is believed that for singularity-free theories, extra degrees of freedom should
possibly manifest themselves in the regime of strong gravity. An ubiquitous statement in this context is that possible quantum
effects should smoothen out classical singularities, but it is fair to say that we are quite far from reaching a consistent quantum
theory of gravity. Now, in the absence of a well established quantum version of GR, a natural alternative is to construct a classical theory 
of gravity itself that is free of singularities. The astrophysical consequences of such a recently discovered theory of modified gravity -- called the 
Eddington inspired Born-Infeld (hereafter EiBI) gravity -- will be the main focus of this paper. In particular, we demonstrate the dynamics of tidal disruption of white dwarfs (WDs) whose interiors are modelled by the Newtonian limit of EiBI gravity, due to an intermediate-mass black hole (IMBH). 
This is done by numerically incorporating effects of modified gravity in smoothed particle hydrodynamics (SPH). 

As we have mentioned, any modification to GR, at cosmological or Planck scales, is associated with incorporating extra degrees of freedom in the theory. 
These are naturally associated with extra parameters that needs to be introduced, the simplest case being a one-parameter modification
of GR which occurs in EiBI gravity. Now, any such modification has its imprints in the low energy 
Newtonian limit, and hence will have observationally important consequences. Indeed, there is a sizeable body of literature by now that seeks to 
constrain these parameters from astrophysical observations (for recent reviews see \cite{SaksteinRev, OlmoRev1}). In this paper,
we study a somewhat different perspective. Namely, assuming the established constraints on EiBI theory, we seek to understand
real time stellar dynamics within the ambits of such constraints and establish how physical quantities are modified due to the freedom in
choosing an extra tuneable parameter in the theory. Importantly, it allows us to compute such physical quantities as a function of the modified gravity parameter so that we can quantify any deviation from their expected GR values. 
Any observable difference of measurable astrophysical quantities from the standard Newtonian case should point to a possibility of  
a modification of GR in general. Conversely, if such features are not observed observationally, it rules out the possible modification
being considered. 

To set the stage and to establish the mathematical formalism used in the rest of the paper, let us first briefly review the
construction of EiBI gravity, that attempts to classically regularise singularities in GR. To this end, recall that 
attempts at regularising singularities have a long history, starting from the celebrated work of 
Born and Infeld \cite{Born1,Born2,BI1}, who constructed a version of electromagnetic theory free from the divergences 
associated with the Maxwell formalism. 
Applied to GR, one such attempt in the recent past is the construction of the theory by Banados and Ferreira (hereafter BF) \cite{BF}, and
builds upon that of \cite{DG} and \cite{Vollick1, Vollick2}. Let us briefly review this construction. Recall that the standard version 
of Born-Infeld electrodynamics replaces the Maxwell Lagrangian ${\mathcal L} = -F^2/4$ by 
${\mathcal L} = b^2(1-\sqrt{|{\rm det}(\eta + F/b})|$, where $F$ is the Maxwell field tensor, $\eta$ the Minkowski metric, and 
$b$ is the Born-Infeld parameter. This theory in principle eliminates the infinite self energy associated with a point particle in Maxwell's theory, 
and the Maxwell Lagrangian is recovered for $F \ll b$. In a similar spirit, Eddington \cite{Eddington, Schrodinger} proposed 
a variant of the Einstein-Hilbert action for GR, and in this formalism, the Lagrangian (apart from pre-factors) 
is taken to be $\sqrt{|{\rm det} R_{\mu\nu}|}$, with $R_{\mu\nu}$ being the Ricci tensor. 
Here, the affine connection is considered as a dynamical variable (the so called Palatini formalism), and variation of the action 
with respect to the connection gives Einstein's equation in the presence of a cosmological constant. Note that 
this latter equation can alternatively be obtained from a variation of the Einstein-Hilbert Lagrangian proportional to 
$\sqrt{|{\rm det}g_{\mu\nu}|}(R - 2\Lambda)$, with the metric $g_{\mu\nu}$ being the dynamical variable, and where 
$R = g^{\mu\nu}R_{\mu\nu}$ being the Ricci scalar and $\Lambda$ is the cosmological constant. 

This line of reasoning was revived in the late 1990s by \cite{DG} who considered a gravitational action of the 
square root form, but with an additional tensor field that needed to be tuned order by order to remove
ghost instabilities. Motivated by \cite{DG} and subsequent works of \cite{Vollick1, Vollick2} which used the Palatini formalism, 
BF considered a Born-Infeld type of action, with a minimal coupling of gravity with matter
fields. The BF Lagrangian reads ${\mathcal L} \sim \sqrt{|{\rm det}[g_{\mu\nu} + \epsilon R_{(\mu\nu)}]|}-\lambda
\sqrt{|{\rm det}g_{\mu\nu}|}$, apart from the matter contribution, and one considers the 
symmetric part of the Ricci tensor in the Lagrangian, denoted by the braces (see e.g., \cite{new4}). In this formalism, $1/\epsilon$ is the 
Born-Infeld mass $M_{BI} \ll M_{Pl}$, the Planck mass. Also, $\lambda$ is a dimensionless non-zero parameter,
related to the cosmological constant, with $\lambda = 1$ giving asymptotically flat solutions. 
The BF formalism leads to singularity-free cosmology, and it was shown by \cite{Pani1} that this avoids singularities that arise 
in gravitational collapse (see also \cite{Delsate1}). The BF modification to GR has come to be known in the literature as EiBI gravity.  

In the non-relativistic limit, EiBI gravity gives rise to a modified Poisson equation, with the modification 
of the low energy limit of Einstein gravity being
characterised by a coupling term that is non-zero only in the presence of matter. Since the Poisson equation
is used as a basic input in many formulas of stellar observables, it is then natural that EiBI theories can have important
consequences in stellar astrophysics. Indeed, there has been a variety of works in the recent past in this direction. The work of \cite{Casanellas1}
proposed tests for the theory using solar constraints. Further, \cite{Avelino1} studied such constraints using cosmological
and astrophysical scenarios, and \cite{Avelino2} obtained bounds on EiBI theories by demanding that electromagnetic
forces dominate gravitational ones in nuclear reaction. More recently, the work of \cite{Banerjee2017} has put constraints on the theory
from an analysis of WDs, and \cite{Olmo1} studied gravitational waves in non-singular EiBI cosmological models. 
Further recent studies on constraining EiBI theories appear in \cite{Pani2012, new1, new5, Rosyadi2019, new2, new3, Banerjee2022}.  
A recent comprehensive review of EiBI gravity and related phenomenological tests appear in the work of \cite{OlmoRev}. 

To be specific, in the low energy limit, EiBI gravity introduces a correction term inside a 
matter source, while GR is recovered outside matter. In the Newtonian limit, the modified Poisson equation takes the form
\begin{equation}
	\nabla^2 \Phi = 4 \pi G \rho + \frac{\kappa}{4}\nabla^2\rho~,
	\label{mod_poi}
\end{equation}
where $\Phi$ represents the gravitational potential, $\rho$ is the matter density, and $G$ denotes the gravitational constant. 
The parameter $\kappa$ represents the correction term introduced by the EiBI theory. Equation (\ref{mod_poi}) is attractive from a numerical 
point of view. Namely, it does not require us to assume spherical symmetry which is not the case in other known examples of modified
gravity theories such as the well studied beyond-Horndeski class of theories (see e.g., \cite{KobayashiRev}). In fact, the form of 
Equation (\ref{mod_poi}) allows us to work in Cartesian coordinates, which is a big advantage over this latter class of theories, as 
for a tidally disrupted star, spherical symmetry no longer holds. We however, need to first construct a spherically
symmetric star within the realms of EiBI gravity before it is studied in presence of tidal fields to simulate its tidal disruption event.

In order to numerically simulate tidal disruption events, we first need to prepare a spherical WD which is in equilibrium. 
In order to do this, we first note that EiBI gravity, the radial acceleration equation in the spherically
symmetric case can be expressed as
\begin{equation}
	a(r) = -\frac{G m(r)}{r^2} - \frac{\kappa}{4}\frac{d\rho}{dr}~,
	\label{acc_eq}
\end{equation}
where $m(r)$ represents the enclosed mass within a radius $r$. Then, the hydrostatic equilibrium 
equation can be derived by considering the balance between gravitational forces and pressure gradients 
in the system. Starting from the modified Poisson equation (Equation (\ref{mod_poi})), we derive the hydrostatic 
equilibrium equation for a spherical star as 
\begin{equation}
	\frac{dP}{dr} = -\frac{G m(r) \rho}{r^2} - \frac{\kappa}{4}\rho \frac{d\rho}{dr}~.
	\label{hydro_eq}
\end{equation}
This equation allows us to study the equilibrium state of the system, incorporating the EiBI correction on the pressure distribution.

The presence of an additional term in the hydrostatic equilibrium equation introduces modifications to the interiors of spherical stars in 
comparison to GR. From Equation (\ref{hydro_eq}), we see that since $d\rho/dr$ is always negative, a positive (negative) value of
$\kappa$ effectively weakens (strengthens) gravity inside a stellar object. This changes the mass-radius relation of WDs. 
The fact that, for a WD of given mass, EiBI gravity changes its compactness by changing its radius is of primary importance in tidal disruption event. For a WD to be tidally disrupted by a black hole without being captured as a whole, its tidal radius $r_t$, estimated by the equation (see \cite{Hills1975}),
\begin{equation}
	r_t \approx \left(\frac{M}{M_{\text{wd}}}\right)^{1/3} R_{\text{wd}},
	\label{tidal_eq}
\end{equation} 
must lie outside the event horizon of the black hole, or to be more precise, its innermost stable orbit. 
Here $M_{\text{wd}}$ and $R_{\text{wd}}$ represent the mass and radius of the WD, respectively, and $M$ represents the mass of the black hole. 
Note that this is an useful but approximate formula and does not take into account the hydrodynamics of the stellar interior. 
It is however important to note that the tidal radius is proportional to $M^{1/3}$, while the event horizon radius is proportional to $M$. As a result, there 
exists a maximum limit for the mass of the black hole, beyond which no disruption occurs, and the WD is captured as a whole, 
without undergoing disruption. 
This limiting black hole mass lies below the supermassive black hole range, namely the intermediate-mass black holes ($\mathcal{O}(10^2)-\mathcal{O}(10^5)$) \cite{Maguire2020}.

Now, from Equation (\ref{tidal_eq}) and our earlier discussion, it follows that for a WD of a given mass, its radius $R_{\text{wd}}$ changes 
due to the effect of EiBI gravity, and thus $r_t$ changes. Hence, modified gravity has a direct impact in tidal disruption events.  
This makes tidal disruption events an excellent tool for testing the effects arising from EiBI gravity. Tidal disruption of WDs in particular, serve as a crucial astrophysical phenomenon for investigating EiBI theory. Their well-known chemical composition, thermodynamic properties, 
and well-understood behaviour in the weak field regime of GR make them ideally suited for performing precise tests of modified gravity
via tidal disruption events. 

Once spherical WDs are prepared after incorporating modified gravity, their dynamics can be studied, including the effects of 
tidal disruption due to an IMBH. In the context of tidal disruption events involving WDs, IMBH holds particular significance. 
In the following sections of the paper, we will set up the framework for studying tidal disruption events and analyze the associated observables. 
Specifically, we will investigate how these observables are influenced by the modified gravity parameter $\kappa$.
The paper is structured as follows: In Section \ref{sec2}, we provide an overview of WD physics and its extension to incorporate EiBI gravity. 
Section \ref{sec3} outlines the methodology for simulating tidal encounters between a WD, modeled with EiBI gravity, and an IMBH using SPH code. 
The simulation results and analysis are presented in Section \ref{sec4}. Finally, Section \ref{sec5} concludes with a summary, discussion on the 
significance of the study, and future prospects.

\section{Modeling WDs with EiBI gravity }
\label{sec2}

WDs represent the end stage of the evolution of low and intermediate-mass stars with masses ranging from  
$\sim 0.1-8$ $\text{M}_{\odot}$. During the evolution of these stars, they consume fuel in their core, lose energy, and shrink, 
while the outer layers of the star expand and the star becomes a red giant. When the outer layers of a red giant is released into space, the hot and dense 
core of the star, primarily composed of carbon and oxygen, remains behind, eventually becoming a WD. These WDs are 
dense, with densities ($\sim 10^6$ $\text{g}\,\text{cm}^{-3}$), and have masses comparable to that of the Sun, but compressed into a volume 
roughly the size of the Earth. The strong gravity of the WD is balanced by the pressure of the degenerate electron gas. We refer the reader to the 
reviews \cite{Koester1990, Isern2022, Fontaine2013, shapiro} for detailed discussions on WDs.

To study the properties of carbon-oxygen WDs, we employ a model based on the formalism developed by  \cite{shapiro, Jain}, 
which we extend to incorporate the EiBI theory, following \cite{Banerjee2017}. 
In this model, we neglect electrostatic interactions and work in the Newtonian limit. We assume a WD in a completely ionized 
state with degenerate electrons at zero temperature. Our goal is to derive an equation of state (EOS) that relates the pressure and density 
within a WD. 

We begin with the fact that the number density of degenerate electrons is given by $n_e = m_e^3 c^3 x^3/(3\pi^2 \hbar^3)$, 
where the `relativity parameter' (dimentionless Fermi momentum) is $x = p_F/(m_e c)$ (not to be confused with the Cartesian coordinate $x$ introduced later)
with $p_F$ being the Fermi momentum, $m_e$ the electron mass, $c$ the speed 
of light, and $\hbar$ is the reduced Planck constant. The total density is the sum of the densities of electrons and carbon ions, 
$\rho = \rho_C + \rho_e$, and is dominated by $\rho_C$ due to the negligible mass of the electron as compared to the carbon atom. This $\rho_C$ is related to the number 
density of the electrons by the relation $\rho_C = m_C n_e/6$, with $m_C$ being the mass of ionized carbon (6 being the atomic number of a 
carbon atom). Thus, the total density is related to the relativity parameter through the relation
\begin{equation}
	\rho \approx m_C m_e^3 c^3 x^3/(18 \pi^2 \hbar^3) = 1.9479 \times 10^9 x^3 \, \text{kg}\,\text{m}^{-3}~.
	\label{rho_eq}
\end{equation}
As the pressure due to non-relativistic carbon ions is much smaller than the pressure contribution from the relativistic electrons, 
so the total pressure is $P \approx P_e$ (the Chandrasekhar approximation, see, e.g., \cite{Boshkayev}). 
The degenerate pressure of the electrons is calculated using the kinetic theory of gases, and is given as
\begin{equation}
	P \approx m_e^4 c^5 \phi(x)/\hbar^3 = 1.4218\times10^{24}\phi(x) \, \text{N}\,\text{m}^{-2}~,
	\label{pres_eq}
\end{equation}
where
\begin{equation}
	\phi(x) = \frac{1}{8\pi^2}\Bigl[x(1+x^2)^{1/2}(2x^2/3-1)+ \log_{e}\bigl[x+(1+x^2)^{1/2}\bigr] \Bigr]~.
	\label{phi_eq}
\end{equation}
Thus, we obtain an EOS for the WD that relates pressure and density through the parameter $x$. 
Inserting the expressions of density and pressure into the mass continuity equation, $dm(r)/dr = 4\pi r^2 \rho$, and Equation (\ref{hydro_eq}) 
results in two first-order coupled differential equations given by
\begin{align}
	\frac{dm}{dr} & = \frac{2 m_C m_e^3 c^3 r^2 x(r)^3}{9\pi \hbar^3} \label{dmdr_eq} \\
	\frac{dx}{dr} & = -\frac{24 \pi ^2 G \hbar^3 m_C m(r) \sqrt{1+x(r)^2}}{r^2\Bigl[144\pi^2 \hbar^3 c^2 m_e x(r) + 
	\kappa c^3 m_C^2 m_e^3 x(r)^2 \sqrt{1+x(r)^2} \Bigr]}~.
	\label{dxdr_eq}
\end{align}
These equations can be solved numerically with initial conditions $m(0) = 0$, and $x(0) = x_0$, where $x_0$ is related to the 
central density. The radius of the star can be calculated using the condition that the pressure at the surface of the star vanishes, 
$P[x(R)] = 0$, which implies $x(R) = 0$ and the total mass of the star can be obtained by $M = m(R)$. From the above equations, 
it is evident that in the EiBI theory, the mass and radius of the star depend on both $x_0$ and $\kappa$, whereas in GR, they depend only on $x_0$. 

In \cite{Casanellas1}, an upper bound on the modified parameter is obtained using solar constraints, resulting in $\kappa \lesssim 3\times 10^5 \, \text{m}^5\text{kg}^{-1}\text{s}^{-2}$.
In the context of neutron stars, \cite{Pani2012} found that $\kappa < 10^{-2} \, \text{m}^5\text{kg}^{-1}\text{s}^{-2}$. Considering cosmological and 
astrophysical scenarios, \cite{Avelino1} derives the bound $\kappa \lesssim GR^2$, where $R$ is related to the Hubble radius or the radius of the compact 
object. In \cite{Rosyadi2019}, the analysis of brown dwarf mass and radii yields $-1.51 \times 10^2 \, \text{m}^5\text{kg}^{-1}\text{s}^{-2} < \kappa < 0.81 \times 
10^2 \, \text{m}^5\text{kg}^{-1}\text{s}^{-2}$ at the $1\sigma$ confidence level and $-1.59 \times 10^2 \, \text{m}^5\text{kg}^{-1}\text{s}^{-2} < \kappa < 1.16 \times 10^2 \, 
\text{m}^5\text{kg}^{-1}\text{s}^{-2}$ at the $5\sigma$ confidence level. By utilizing the mass-radius data of cataclysmic variables, \cite{Banerjee2022} derives the 
bounds $0.005 \leq \kappa/GR_{\odot}^2 \leq 0.352$ at the $1\sigma$ level and $-0.315 \leq \kappa/GR_{\odot}^2 \leq 0.597$ at the $5\sigma$ level. 

Important in our context will be the study by \cite{Banerjee2017}, where $\kappa$ is constrained through a $\chi^2$ analysis of observational data from twelve 
WDs \cite{Holberg2012}. The results yield a bound of $-0.7 \times 10^3 \, \text{m}^5\text{kg}^{-1}\text{s}^{-2} < \kappa < 1.66 \times 10^3 \, 
\text{m}^5\text{kg}^{-1}\text{s}^{-2}$ at the $1\sigma$ confidence level and $-1.598 \times 10^3 \, \text{m}^5\text{kg}^{-1}\text{s}^{-2} < \kappa < 4.858 \times 
10^3 \, \text{m}^5\text{kg}^{-1}\text{s}^{-2}$ at the $5\sigma$ confidence level. These WDs, with mass range $\sim 0.5-1 M_{\odot}$ 
in the left panel of Figure \ref{fig.den} make these objects a viable set for our study in this paper, as the error bars in their mass measurements can
be attributed to the presence of modified gravity. We should also point out that the same study 
reports that when considering super-Chandrasekhar white dwarfs with masses up to $2.8 M_{\odot}$ \cite{Taubenberger}, $\kappa < 0.35 \times 10^2 \, 
\text{m}^5\text{kg}^{-1}\text{s}^{-2}$ ($5\sigma$). These will however be excluded here in the absence of a well known equation of state and possible 
effects of magnetic fields in such exotic stars. 
In our study, we thus focus on WDs in the mass range $\sim 0.5-1.0 M_{\odot}$, utilizing the $5\sigma$ bound obtained by \cite{Banerjee2017} mentioned above. We employ these 
established bounds on the modified gravity parameter, $\kappa$, to investigate the observational effects of EiBI theory on WDs, using the physics of tidal disruptions.

\section{Formalism and methodology}
\label{sec3}

In this section, we describe the formalism and the methodology employed to investigate the effects of EiBI gravity on tidal disruption events of 
WDs by an IMBH. Tidal disruption events occur when a star comes close to a black 
hole and experiences non-local disruptive forces due to the strong gravitational field. Our objective is to analyze the observational 
signatures of tidal disruption events in the presence of EiBI gravity. To achieve this, we performed three-dimensional hydrodynamical 
simulations based on SPH. The reader is referred to \cite{Banerjee2023} 
for a detailed description of the numerical methods and the code employed to simulate tidal disruption events.

\subsection{Hydrodynamics}

SPH is a Lagrangian method that models fluid stars via a set of particles. In this method, the fluid properties, such as density, pressure, and velocity, are calculated for each particle. These properties are `smoothed' using a 
fixed number of neighbouring particles using a M6 quintic spline kernel. The forces acting on each particle are determined through a 
binary tree algorithm, which employs a tree opening angle of $\theta = 0.5$ to restrict the number of neighbouring particles taken into account. 
To account for the dissipation of energy due to the viscosity of the fluid, artificial viscosity is introduced with standard artificial viscosity 
parameters, $\alpha^{\text{AV}} = 1.0$ and $\beta^{\text{AV}} = 2.0$. To calculate the external gravitational force exerted on each 
particle by the black hole, we followed the same approach as in \cite{Banerjee2023}, in which each particle 
experiences the relativistic acceleration in Schwarzschild space-time. This approach takes into account the effects of general relativity 
and is therefore more accurate than Newtonian gravity when modelling tidal forces. Finally, the SPH equations are evolved using the leapfrog approach at each time step, and a global time step is employed to ensure numerical stability. 

\subsection{White dwarf EOS in SPH}

To incorporate the zero-temperature equation of state for the electron gas in SPH, we follow a few key steps. First, we 
estimate the density of each particle using a kernel function that assigns weights to neighbouring particles based on their distances. 
Next, we compute the relativity parameter, $x$ for each particle from density using Equation (\ref{rho_eq}). Once we have the 
relativity parameter for each particle, the pressure is calculated using Equation (\ref{pres_eq}) and Equation (\ref{phi_eq}) for each particle. Taking into account the zero-temperature equation of state, the sound speed is updated, and its relationship to the relativity parameter is given by
\begin{align}
	c_s^2 & = \frac{\partial P}{\partial \rho} = \frac{2 m_e c^2}{m_C}	\frac{x^2}{\sqrt{1+x^2}} \notag\\
	      & = 8.2173 \times 10^{12} 	\frac{x^2}{\sqrt{1+x^2}} \, \text{m}^2\,\text{s}^{-2}~.
\end{align}
This sound speed, determined by the above equation, is crucial for accurately capturing the dynamics in SPH simulations.

Finally, the computed pressure values are then utilized to determine the forces acting on each particle over time. The above EOS is derived with the assumption that the degeneracy pressure is significantly higher than the thermal pressure of the gas. 
This is justified due to the fact that during tidal disruption, the high compression of matter leads to high densities and low temperatures. 
At these low temperatures, most of the electrons are in their lowest energy state and degeneracy pressure dominates 
over thermal pressure.

\subsection{Implementation of EiBI gravity}

In SPH, self-gravitational forces can be calculated using the near-field and the far-field approaches. 
In the near-field approach, the gravitational force on each particle is determined by summing over the contributions from its 
neighbouring particles within a certain smoothing length. The gravitational softening kernel, which is based on the distance 
between particles and the smoothing length, is used to weight the contributions from each neighbour. In contrast, the far-field approach 
calculates the gravitational force on a particle due to a group of distant particles using the multiple moment expansion. 

To incorporate the EiBI theory into the SPH framework, it is necessary to use the modified Poisson equation 
given in Equation (\ref{mod_poi}). This equation directly affects the calculation of self-gravitational forces between particles.
The modified Poisson equation introduces an extra term $\frac{\kappa}{4}\nabla^2\rho$, which modifies the gravitational softening 
kernel within the smoothing length. This modified kernel captures the gravitational interaction within the vicinity of a particle, 
providing a more accurate representation of the EiBI gravity effects.
However, outside the smoothing length, the gravitational softening kernel remains the same as in standard gravity, and is proportional 
to $1/r^2$. This is because beyond the smoothing radius, particles do not contribute to the density and hydrodynamic force on the particle of interest. Therefore, these particles are not considered a part of the fluid element of interest. As we know that the EiBI modification is prevalent only within the matter source (in this case the fluid element), only the near field gravity is modified, whereas in vacuum, EiBI tends to GR, so the far field gravity is calculated using the Newtonian gravity without any modification. 

Now, the modified gravitational softening kernel is related to the density kernel using the modified Poisson equation, given by
\begin{equation}
	W(r,h) = \frac{1}{4\pi r^2}\Biggl[\frac{\partial}{\partial r}\left(r^2\frac{\partial\phi(r,h)}{\partial r}\right) - 
	\kappa'\frac{\partial}{\partial r}\left(r^2\frac{\partial W(r,h)}{\partial r} \right)\Biggr]~,
\end{equation}
where $W(r,h)$ and $\phi(r,h)$ are the density and gravitational softening kernel respectively and $\kappa' = \kappa/(4 G)$. 
By integrating the above equation, we can obtain the derivative of the softening kernel, $\partial\phi/\partial r$, which is given by
\begin{equation}
	\frac{\partial\phi(r,h)}{\partial r} = \frac{4\pi}{r^2} \int_{}^{r} r'^2 W(r')\,dr'+ \kappa'\frac{\partial W(r,h)}{\partial r} + \frac{C_1}{r^2}~.
\end{equation}
The constant $C_1$ is determined by imposing the condition that the standard Newtonian inverse square law is recovered 
beyond the smoothing length of the kernel.

Further integrating the equation for ${\partial\phi(r,h)}/{\partial r}$ yields the softening kernel, expressed as
\begin{equation}
	\phi(r,h) = \int_{}^{r} \left(\frac{4\pi}{\tilde{r}^2} \int_{}^{\tilde{r}} r'^2 W(r')dr'\right)d\tilde{r} + \kappa' W(r,h) - \frac{C_1}{r} + C_2~.
\end{equation}
Here, the constant $C_2$ is determined by considering the asymptotic behaviour ($\phi \to 0 \ \text{as} \ r \to \infty$) of the softening kernel.
The incorporation of EiBI theory into the SPH framework allows an accurate representation of the effects of EiBI gravity within a star. 
This inclusion has significant implications for the dynamics of tidal disruption events, which will be explored in the subsequent section.

In Appendix \ref{appA}, we present the analytical forms of the softening kernel, $\phi(r,h)$, and the derivative of the softening kernel
$\partial\phi/\partial r$, which are used in our simulations.

\subsection{Initial density profile}

\begin{figure}[H]
	\centering 
	\includegraphics[scale=0.295]{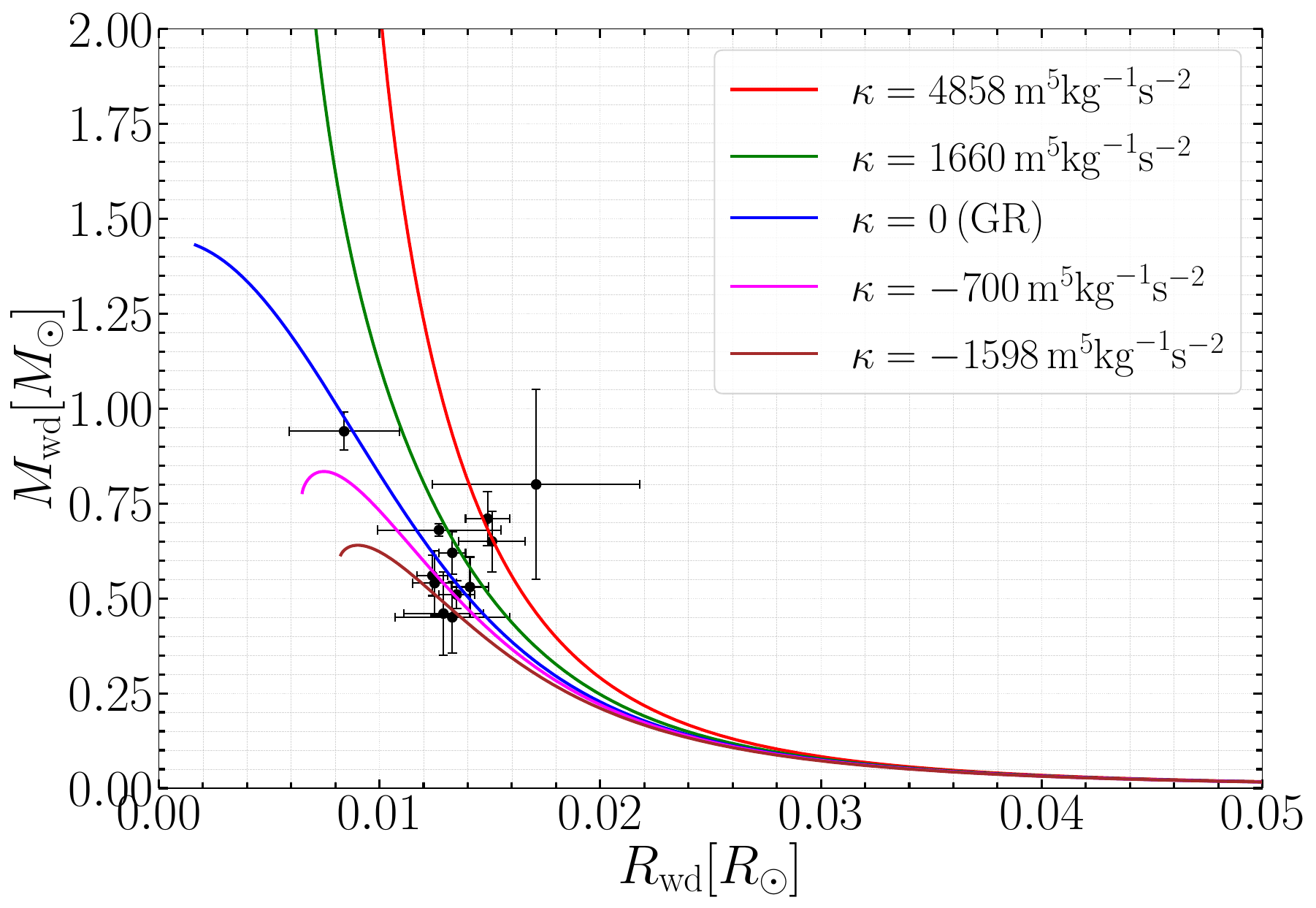}
	\includegraphics[scale=0.295]{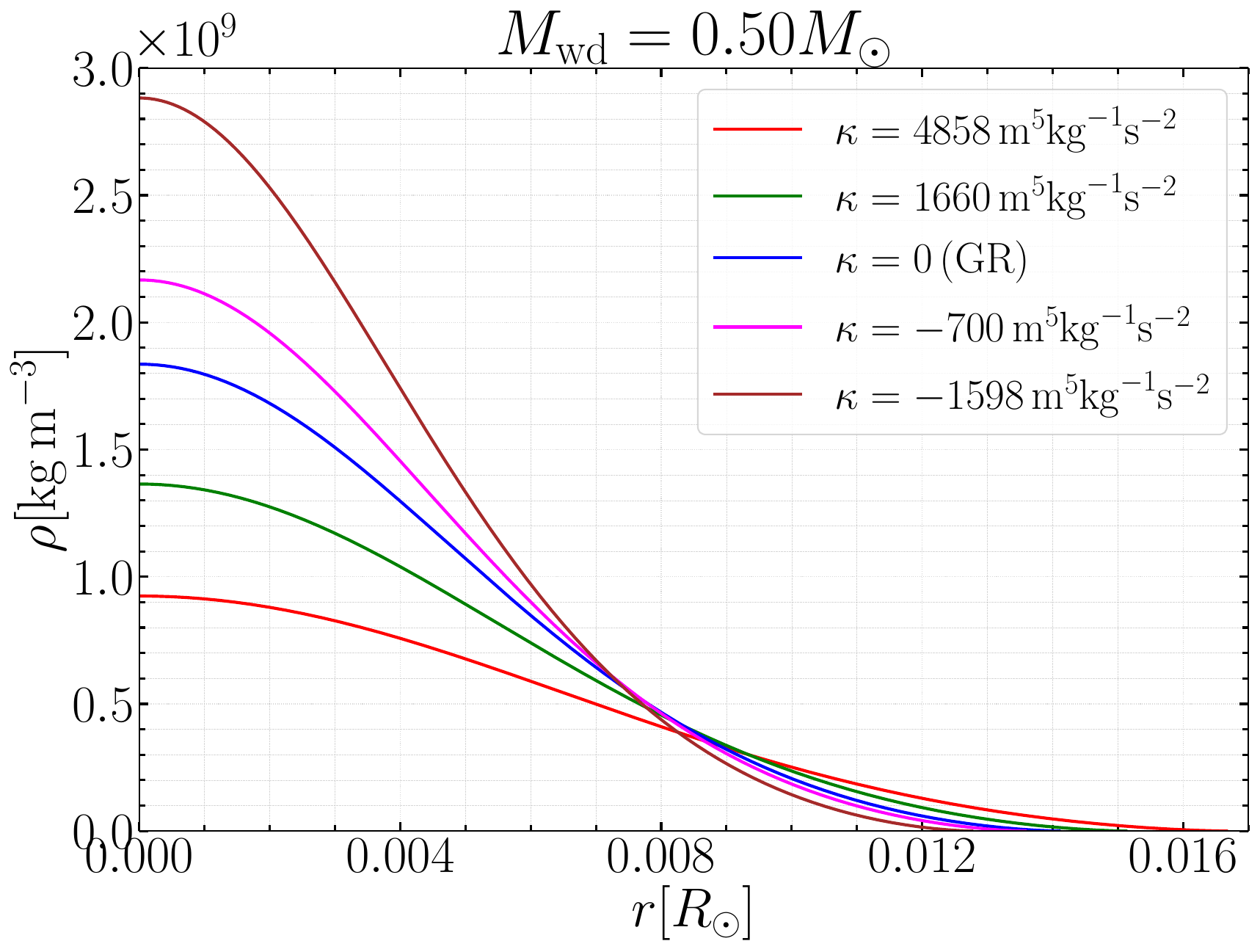}
	\caption{\textbf{Left panel:} The mass-radius relations of WDs are presented for different $\kappa$ values in EiBI gravity. Observed mass-radius of WDs from \cite{Holberg2012}, along with their associated error bars, are represented by black dots.
	\textbf{Right panel:} Radial density profile for the five WDs with different $\kappa$ values of mass $0.50 \, \text{M}_\odot$.}
	\label{fig.den}
\end{figure}

In order to obtain the initial density profile, we employ a numerical solution of Equations (\ref{dmdr_eq}) and (\ref{dxdr_eq}) as discussed 
in Section \ref{sec2}. These equations yield the mass-radius relation for a chosen value of the parameter $\kappa$. 
The influence of $\kappa$ on the mass-radius relationship is shown in Figure \ref{fig.den} (left panel), where we present 
the mass-radius relations for various $\kappa$ values. As discussed in the introduction, for $\kappa > 0$, the additional term counteracts 
the self-gravitational term, allowing the WD to support more mass. Conversely, for $\kappa < 0$, the opposite behaviour occurs, 
resulting in the WD being able to support less mass. Once the mass and radius for a specific $\kappa$ are determined, 
the density profile of the star is obtained using the profile $x(r)$ derived from Equations (\ref{dmdr_eq}) and (\ref{dxdr_eq}). 
Finally, by using the $x(r)$ profile in Equation (\ref{rho_eq}), we obtain the initial density profile. Figure \ref{fig.den} (right panel) 
displays the radial density profiles for WDs with a mass of $0.50 \, \text{M}_\odot$ for various $\kappa$ values.
After obtaining the initial density profile, it is incorporated into the SPH code. Initially, the particles are placed within a closed-packed sphere, 
which is then stretched using the stretch map technique (see \cite{Herant}) to match the desired density profile. Subsequently, the particle distribution evolves 
in isolation to attain a relaxed configuration. Once the relaxed profiles are obtained, we plot them in Appendix \ref{appB}, alongside 
the initial density profiles generated through the procedure mentioned above.

Furthermore, it is important to note that the central densities involved in our work exceed $10^5 \, \text{g}\,\text{cm}^{-3}$. In a study by \cite{Boshkayev}, it was highlighted that temperature effects become significant when the density drops below $10^5 \, \text{g}\,\text{cm}^{-3}$. Thus, the assumption of neglecting finite temperature effects proves to be a valid approximation for our study.

\subsection{Simulation details}

To investigate the effects of EiBI gravity on tidal disruption observables, we performed 19 simulations of tidal disruption events. 
The central IMBH is modelled as a Schwarzschild black hole with a mass of $M = 10^3 M_\odot$, placed at the origin of the coordinate system. 
The Schwarzschild radius is denoted by $r_s = {{2GM}}/{{c^2}}$. Any particle that crosses this radius is removed from the system. 
We construct relaxed WDs with masses of $0.50, 0.75, 1.00 \, M_\odot$. From Figure \ref{fig.den} (left panel), 
it becomes apparent that the influence of EiBI gravity increases as we move towards higher masses, justifying our choice of WD mass values. It should be noted that decreasing the negative value of $\kappa$ reduces the maximum mass limit.  
As we lower the $\kappa$ value from $-0.7 \times 10^3 \, \text{m}^5\text{kg}^{-1}\text{s}^{-2}$ to $-1.598 \times 
10^3 \, \text{m}^5\text{kg}^{-1}\text{s}^{-2}$, the maximum mass for which a stable WD can exist decreases. 
Consequently, for $\kappa = -0.7 \times 10^3 \, \text{m}^5\text{kg}^{-1}\text{s}^{-2}$ say, there is no WD with a mass of 
$1.00 M_\odot$, while for $\kappa = -1.598 \times 10^3 \, \text{m}^5\text{kg}^{-1}\text{s}^{-2}$, WDs with 
masses of $0.75 M_\odot$ and $1.00 M_\odot$ do not exist.

In this work, we place the relaxed stars in parabolic orbits around the black hole. To ensure a meaningful comparison and to 
isolate the effects of EiBI gravity, we begin by fixing the pericenter distance ($r_p$) from the black hole and passing different white 
dwarfs around it. By doing so, we maintain a constant physical distance, ensuring that the tidal field strength experienced by the various 
WDs at the pericenter remains the same. Consequently, any differences in the observables arise solely from the influence 
of EiBI gravity. In another approach, we fix the impact parameter, $\beta = r_t/r_p$, where $r_t$ represents the tidal radius. This approach allows 
us to fix the average strength of the tidal field experienced by the WDs at the pericenter position relative to the tidal radius.  
In our simulations, we set the initial separation as $350 r_g$ for cases with a fixed pericenter distance and $5r_t$ for cases with a 
constant $\beta$. Here, $r_g=G M/c^2$ represents the gravitational radius. The initial positions and velocities in Cartesian coordinate 
are obtained by the relativistic description given in \cite{Banerjee2023}.

\begin{table}[ht]
	\centering
	\caption{The parameters used in the simulation of Tidal disruption events for fixed pericenter distance ($r_p$).}
	\label{tab1}
	\renewcommand{\arraystretch}{1.2} 
	\begin{tabularx}{\textwidth}{XXXXXX}
		\toprule
		\toprule
		${M}_{\text{wd}}$ [${M}_{\odot}$] & $\kappa$ [$\text{m}^5\text{kg}^{-1}\text{s}^{-2}$] & $R_{\text{wd}}$ [$R_{\odot}$] & $r_p$ [$r_g$] & $r_t$ [$r_g$] & $\beta$ \\
		\midrule
		0.50 & -1598.00 & 0.0128 & 70.00 & 75.88 & 1.084 \\
		0.50 &  -700.00 & 0.0136 & 70.00 & 80.74 & 1.153 \\
		0.50 &     0.00 & 0.0141 & 70.00 & 83.84 & 1.198 \\
		0.50 &  1660.00 & 0.0151 & 70.00 & 89.93 & 1.285 \\
		0.50 &  4858.00 & 0.0167 & 70.00 & 99.10 & 1.416 \\
		\hline
		0.75 &  -700.00 & 0.0098 & 70.00 & 50.69 & 0.724 \\
		0.75 &     0.00 & 0.0109 & 70.00 & 56.64 & 0.809 \\
		0.75 &  1660.00 & 0.0125 & 70.00 & 64.84 & 0.926 \\
		0.75 &  4858.00 & 0.0144 & 70.00 & 74.94 & 1.071 \\
		\hline
		1.00 &  1660.00 & 0.0107 & 70.00 & 50.41 & 0.720 \\
		1.00 &  4858.00 & 0.0130 & 70.00 & 61.29 & 0.876 \\
		\bottomrule
	\end{tabularx}
\end{table}

Tables \ref{tab1} and \ref{tab2} present the parameter space for our tidal disruption simulations. Both tables provide information on the mass, radius, 
$\kappa$ value, and tidal radius of each WD. Here, we calculate the tidal radii of WDs by using Equation (\ref{tidal_eq}). In Table \ref{tab1}, 
we maintain a fixed pericenter distance of $r_p = 70 r_g$. As $r_t$ varies for different WDs having a fixed $r_p$, the $\beta$ values differ for 
different stars. In Table \ref{tab2}, we consider two different values for the impact parameter: $\beta = 0.80$ and $\beta = 1.00$ for a fixed WD 
mass ($M_{\text{wd}} = 0.75 M_{\odot}$). This set of parameter values allows us to study the effect of EiBI gravity in both partial and full disruption scenarios.

\begin{table}[ht]
	\centering
	\caption{The parameters used in the simulation of Tidal disruption events for fixed $\beta$ values.}
	\label{tab2}
	\renewcommand{\arraystretch}{1.2} 
	\begin{tabularx}{\textwidth}{XXXXXX}
		\toprule
		\toprule
		${M}_{\text{wd}}$ [${M}_{\odot}$] & $\kappa$ [$\text{m}^5\text{kg}^{-1}\text{s}^{-2}$] & $R_{\text{wd}}$ [$R_{\odot}$] & $r_t$ [$r_g$] & $\beta$ & $r_p$ [$r_g$]\\
		\midrule
		0.75 &  -700.00 & 0.0098 & 50.69 & 0.80 & 63.37\\
		0.75 &     0.00 & 0.0109 & 56.64 & 0.80 & 70.80\\
		0.75 &  1660.00 & 0.0125 & 64.84 & 0.80 & 81.06\\
		0.75 &  4858.00 & 0.0144 & 74.94 & 0.80 & 93.68\\
		\hline
		0.75 &  -700.00 & 0.0098 & 50.69 & 1.00 & 50.69\\
		0.75 &     0.00 & 0.0109 & 56.64 & 1.00 & 56.64\\
		0.75 &  1660.00 & 0.0125 & 64.84 & 1.00 & 64.84\\
		0.75 &  4858.00 & 0.0144 & 74.94 & 1.00 & 74.94\\
		\bottomrule
	\end{tabularx}
\end{table}

As we set up the star in a trajectory, our goal is to compute tidal disruption observables and find out their dependence on the modified gravity parameter $\kappa$. A key observable of interest is the peak fallback rate, defined as the rate at which the disrupted 
debris falls back towards the pericenter position. We follow \cite{Coughlin2015, Golightly2019, Miles2020} to calculate the peak fallback rate directly from the simulation by capturing the mass accretion towards the pericenter as the disrupted debris falls back. Directly measuring the rate at which debris is accreted onto the black hole allows us to track the fallback rate accurately. This method overcomes the limitations of the frozen-in approximation, which neglects the self-gravity of the debris (see \cite{Coughlin2015, Golightly2019, Cufari2022}). Once the disrupted debris is beyond the pericenter position, they are accreted by the black hole eventually. To ensure an efficient simulation, we increase the accretion radius to $r_{acc} \simeq 3 r_t$ to efficiently remove the debris intended for accretion. Once any bound debris falls back to this accretion radius, it is removed from the system and contributes to the fallback rate. At this point, we need to mention that the fallback rate obtained through this radius may differ from the true accretion rate, which requires modeling of the accretion flow around the black hole and disk formation. However, if the debris accretes onto the black hole rapidly enough and there is no significant delay in the circularization process (\cite{Mockler2019} found that this delay time is very small in observed tidal disruption events), then the computed fallback rates obtained from our simulations closely 
correspond to the true accretion rates. 

Additionally, in partial disruption, outer layers are ejected and a fraction of the star remains 
as a self-bound core. We adopt a methodology similar to the one described in \cite{Guillochon2013}, 
to calculate the self-bound core using an iterative approach based on the particle energy. 
After identifying the core particles, we compute the core properties such as mass, specific energy, specific angular momentum, etc. 
From these core properties, we further compute observables such as kick velocity and trajectory deviations (see \cite{Manukian2013, Gafton2015, Banerjee2023}). 
These could have implications in various areas, including black hole mass determination, understanding hypervelocity stars etc. 

In partial disruption scenarios, the presence of a high-density core leads to extremely small time steps. 
This makes it computationally expensive to simulate the fallback process over longer durations. To address this, once the bound core moves
a significant distance away from the black hole ($> 45r_t$), the core particles are replaced by a sink particle following \cite{PHANTOM}.  
The position and velocity of the sink particle are set to the center of mass position and velocity of all the core particles. The accretion radius of 
the sink particle is equal to the maximum distance of any bound particle from the core's center of mass. We have crosschecked this by introducing 
the sink particle at different distances from the black hole and found no difference in the peak fallback rates. This approach allows us to deal with the 
computational challenges posed by the high-density core effectively and continue the simulations with improved efficiency and accuracy.

We use $5\times 10^5$ number of particles to simulate the stars. Importantly, it should be noted that we performed additional simulations using $1\times10^5$ and $1\times10^6$ particles, and we observe that the results remain consistent across different resolutions.

\section{Results}
\label{sec4}

When a WD approaches a black hole, the gravitational force from the black hole is stronger on the side of the WD closest to it compared to the farther side. This difference in gravitational force deforms the WD, causing it to become elongated in the radial direction while compressed in the vertical and azimuthal directions. As the WD continues to get closer to the black hole, the deformation increases, and the tidal force exerted on the WD increases. At the pericenter, which represents the closest point of approach, the extent of disruption is determined by the impact factor, which is the ratio of tidal radius to the pericenter distance. If the WD enters well within the tidal radius, the tidal forces overcome the WD's self-gravity, leading to full disruption, where the WD is completely torn apart, and its material forms a stream of disrupted debris. On the other hand, if the WD approaches the black hole from a distance well outside the tidal radius, only a portion of the WD torn apart, resulting in partial disruption. In such cases, the central core can either remain bound to the black hole or gain enough energy to escape its gravitational influence, possibly becoming a hypervelocity star. 

Additionally, the disrupted debris from both full and partial disruptions that is bound to the black hole experiences fallback onto the black hole, forming an accretion disk. As the debris circularizes, it releases gravitational potential energy, emitting radiation across various wavelengths. The rate at which the debris falls back, known as the fallback rate, determines the luminosity of the tidal disruption event. The light curve exhibits characteristic features, such as an initial rise in brightness followed by a peak and subsequent fading over time, providing valuable insights into the dynamics and properties of the disrupted WD, the accretion processes, black hole mass, etc.

In the following two subsections, we will discuss the results obtained from our tidal disruption simulations, employing two different approaches: fixing the pericenter distance and fixing the impact parameter. In both of the approaches, we study the observed effects of the modified gravity in partial and full disruption scenarios.

\subsection{Fixed pericenter distance ($r_p$) simulations}
\label{fixedrp}

In this subsection, we focus on the simulations performed with a fixed pericenter distance of $r_p = 70 r_g$, where $r_g$ represents the gravitational radius as mentioned earlier. This approach holds greater observational significance, as it maintains constant physical distances from the black hole for all WDs, ensuring a uniform tidal field strength and effectively isolating the effects of EiBI gravity. As $r_p$ remains constant, we vary the impact parameter, $\beta$ for different stars, and the extent of disruption varies among the stars. Interestingly, during the simulations, we observed a distinct core formation occurring when $\beta \lesssim 0.90$. From Table \ref{tab1}, it is evident that four WDs fall within this category. For these particular stars, partial disruption takes place, leading to the formation of the core with asymmetric tails. This asymmetry arises due to the lower mass ratio $q = M/M_{\text{wd}} \sim 10^3$. 

\begin{figure}[H]
	\centering 
	\begin{subfigure}{0.45\textwidth}
		\centering
		\includegraphics[width=\textwidth]{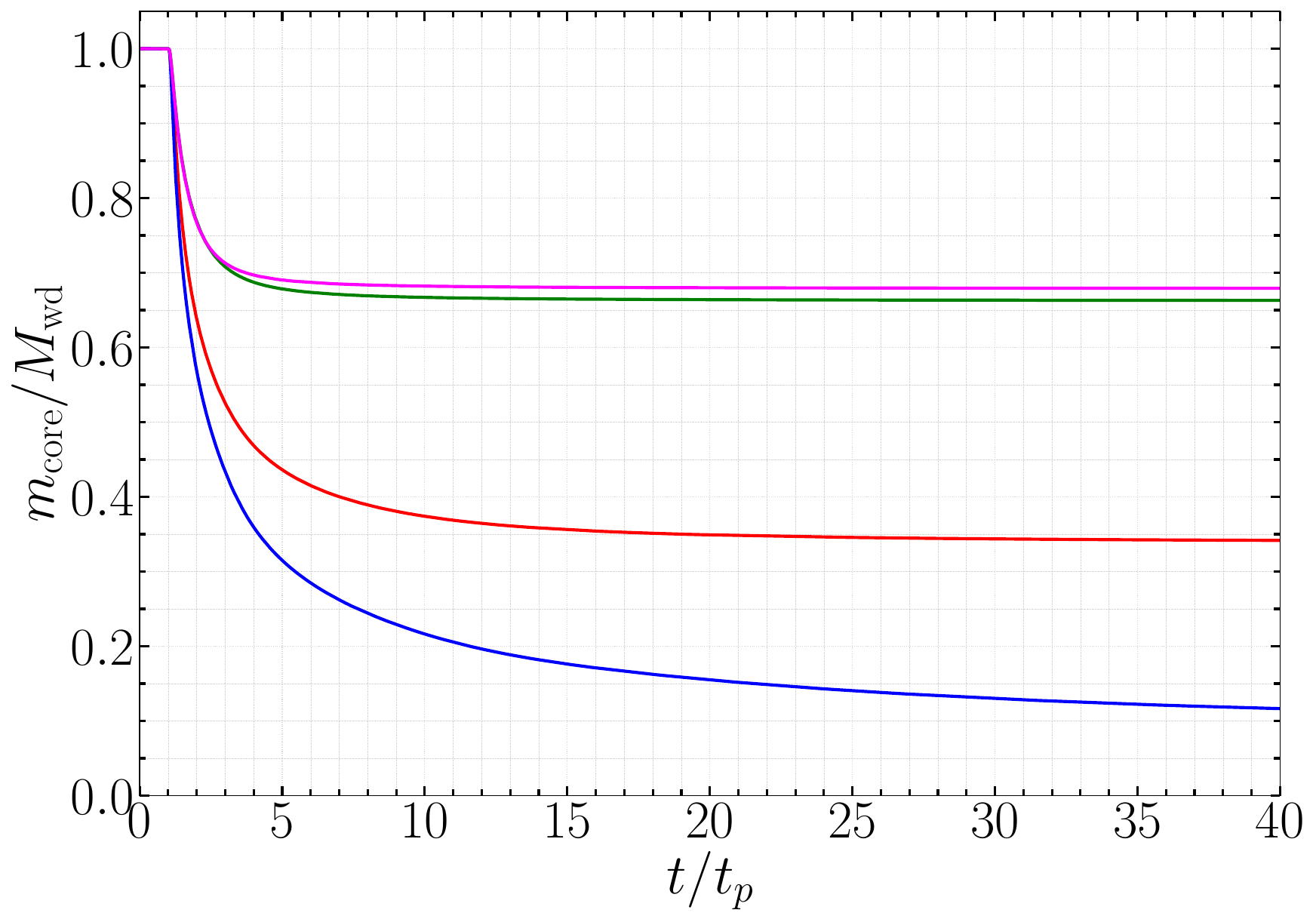}
	\end{subfigure}\hfill
	\begin{subfigure}{0.45\textwidth}
		\centering
		\includegraphics[width=\textwidth]{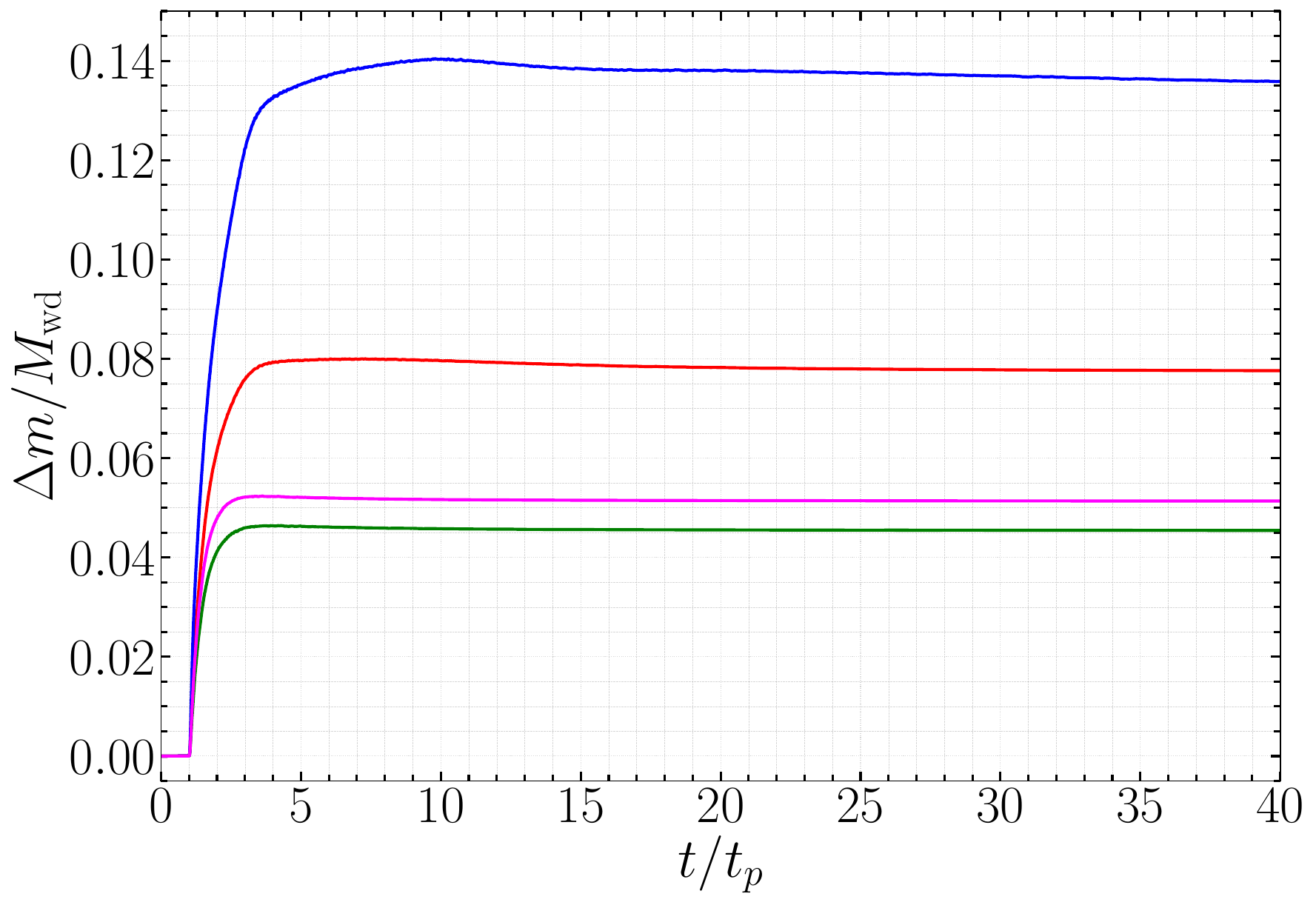}
	\end{subfigure}\\
	\begin{subfigure}{0.45\textwidth}
		\centering
		\includegraphics[width=\textwidth]{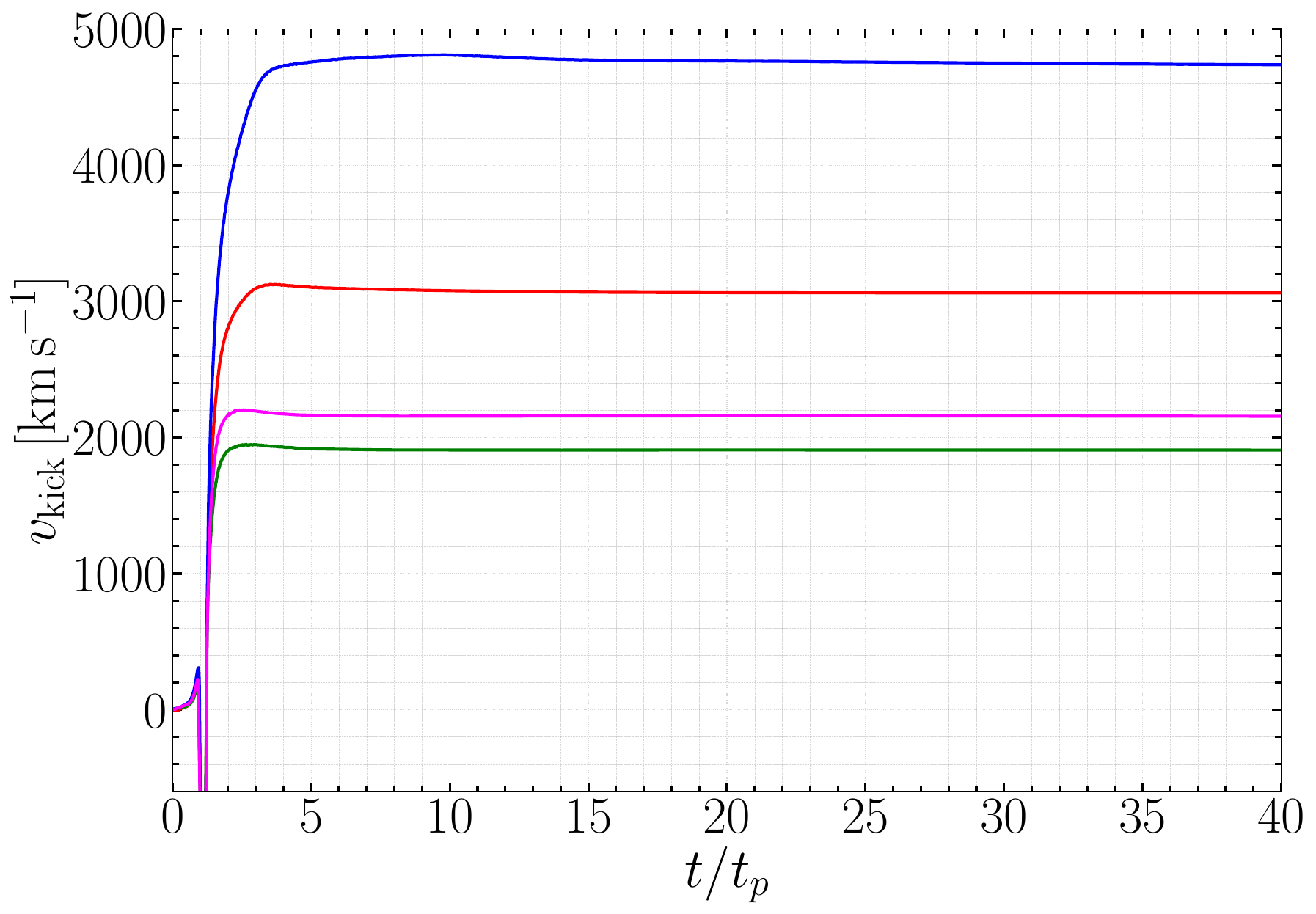}
	\end{subfigure}\hfill
	\begin{subfigure}{0.45\textwidth}
		\centering
		\includegraphics[width=\textwidth]{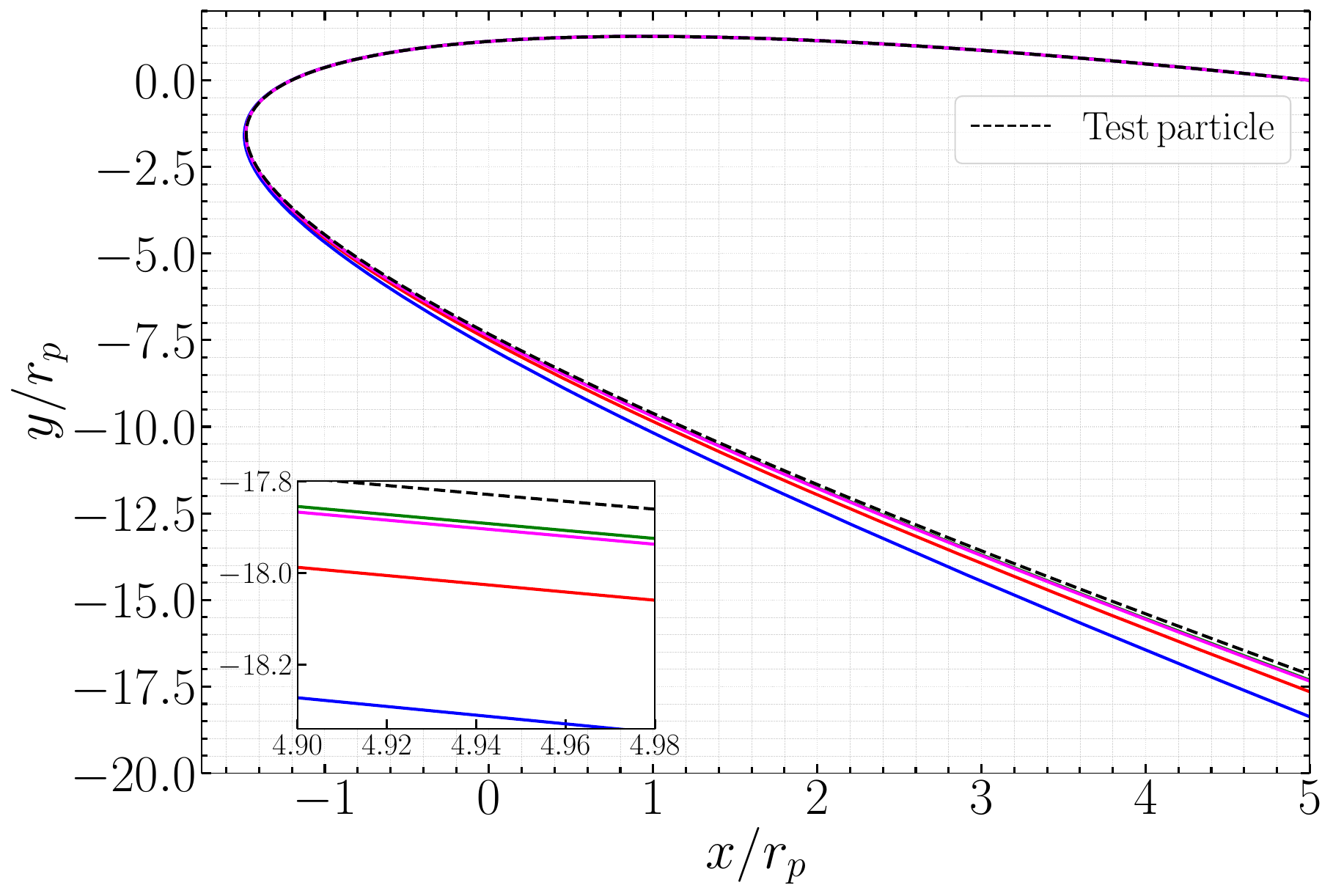}
	\end{subfigure}
	\vspace{0.2cm}
	\includegraphics[scale=0.882]{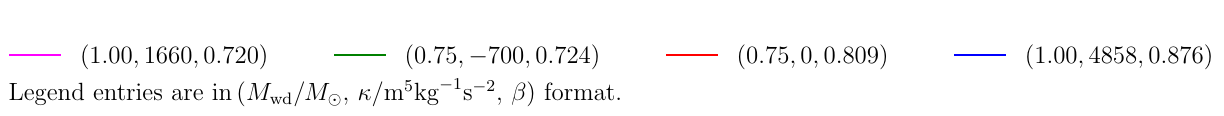}
	\caption{\textbf{Top Left:} Evolution of the bound core mass fraction with normalized time is presented for various WDs mentioned in the legend. \textbf{Top Right:} Variation of the mass difference between two tidal tails with time is shown for various WDs mentioned in the legend. \textbf{Bottom Left:} Plot of the kick velocity of the core over time. \textbf{Bottom Right:} Trajectory deviations for different stars from the initial parabolic trajectory are presented. The inset in the bottom left corner shows a zoomed section of the trajectories after tidal interaction.}
	\label{fig.core_obs}
\end{figure}

In Figure \ref{fig.core_obs}, in the Top Left panel, we present the variations in bound core masses ($m_{\text{core}}$) relative to the initial WD masses ($M_{\text{wd}}$) over time. The time is normalized to the time at which different WDs reach their pericenter positions. As we discussed previously, in the case of partial disruption, the outer layers of the WD are torn apart, leaving behind a self-gravitating core. As a result, the bound core mass fraction gradually decreases from its initial value of 1.0 as the initial WD loses mass. Eventually, the core separates from the tails and the mass fraction stabilizes at a saturated value. To extend the simulations for a longer duration, we replace the core particles with a sink particle to account for the fallback onto the black hole. The figure demonstrates that with increasing $\beta$ values, there is an increase in mass loss. This is because more deep encounters lead to a greater loss of mass from the initial WD.

Moving to the Top Right panel of Figure \ref{fig.core_obs}, we present the mass difference ($\Delta m$) between the two tails relative to the initial WD mass. To understand this behavior, we need to consider the variation of asymmetry with two parameters: $\beta$ and $q$. As $q$ decreases, the difference in the tidal field across the star increases, enhancing asymmetry. Similarly, increasing $\beta$ also contributes to increased asymmetry. These effects are evident in the figure. Among the four WDs, the one with a mass of $1.00 M_{\odot}$ and $\kappa = 4858 \, \text{m}^5\text{kg}^{-1}\text{s}^{-2}$ has the highest $\beta$ value of $0.876$ and a lower $q$ value, resulting in the highest observed asymmetry. However, for the WDs with masses of $1.00 M_{\odot}$ and $\kappa = 1660 \, \text{m}^5\text{kg}^{-1}\text{s}^{-2}$ and $0.75 M_{\odot}$ and $\kappa = -700 \, \text{m}^5\text{kg}^{-1}\text{s}^{-2}$, the $\beta$ values are almost the same. Here, due to the decrease in $q$, the WD with $1.00 M_{\odot}$ exhibits higher asymmetry compared to the $0.75 M_{\odot}$ WD.

In partial disruption, when only a portion of the star is torn apart, an interesting phenomenon occurs due to the conservation of linear momentum. The momentum carried away by the bound tail imparts a `kick' on the remaining self-bound core. As a result, there is an increase in core velocity that translates into the increase in the specific orbital energy and the specific angular momentum of the core. The kick velocity, which quantifies the increase in specific orbital energy of the core, defined as $v_{\text{kick}}= \sqrt{2(\epsilon_{\text{core}}-\epsilon_{\text{in}})}$, 
where $\epsilon_{\text{core}}$ represents the specific orbital energy of the core and $\epsilon_{\text{in}}$ represents the initial specific orbital energy. The variation of kick velocity over time is depicted in the Bottom Left panel of Figure \ref{fig.core_obs}. Asymmetry in the mass loss plays a significant role in increasing the specific orbital energy and, consequently, the kick velocity. From the figure, it is evident that as the asymmetry increases, so does the kick velocity. These kick velocities typically reach values on the order of $\sim 10^3 \text{km}\,\text{s}^{-1}$, a range in which several observed hypervelocity stars fall.
\begin{figure}[H]
	\centering 
	\begin{subfigure}{0.45\textwidth}
		\centering
		\includegraphics[width=\textwidth]{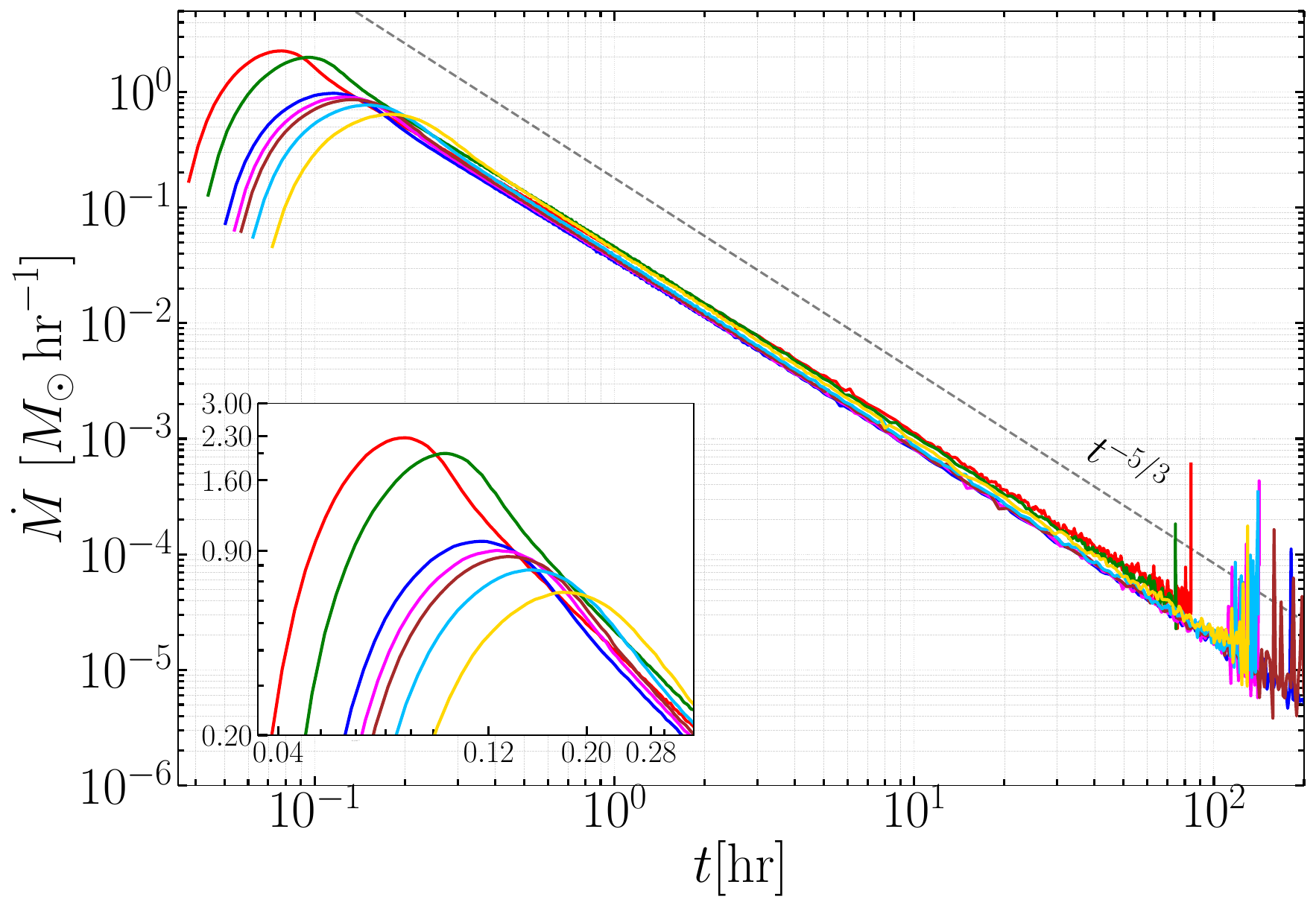}
	\end{subfigure}\hfill
	\begin{subfigure}{0.45\textwidth}
		\centering
		\includegraphics[width=\textwidth]{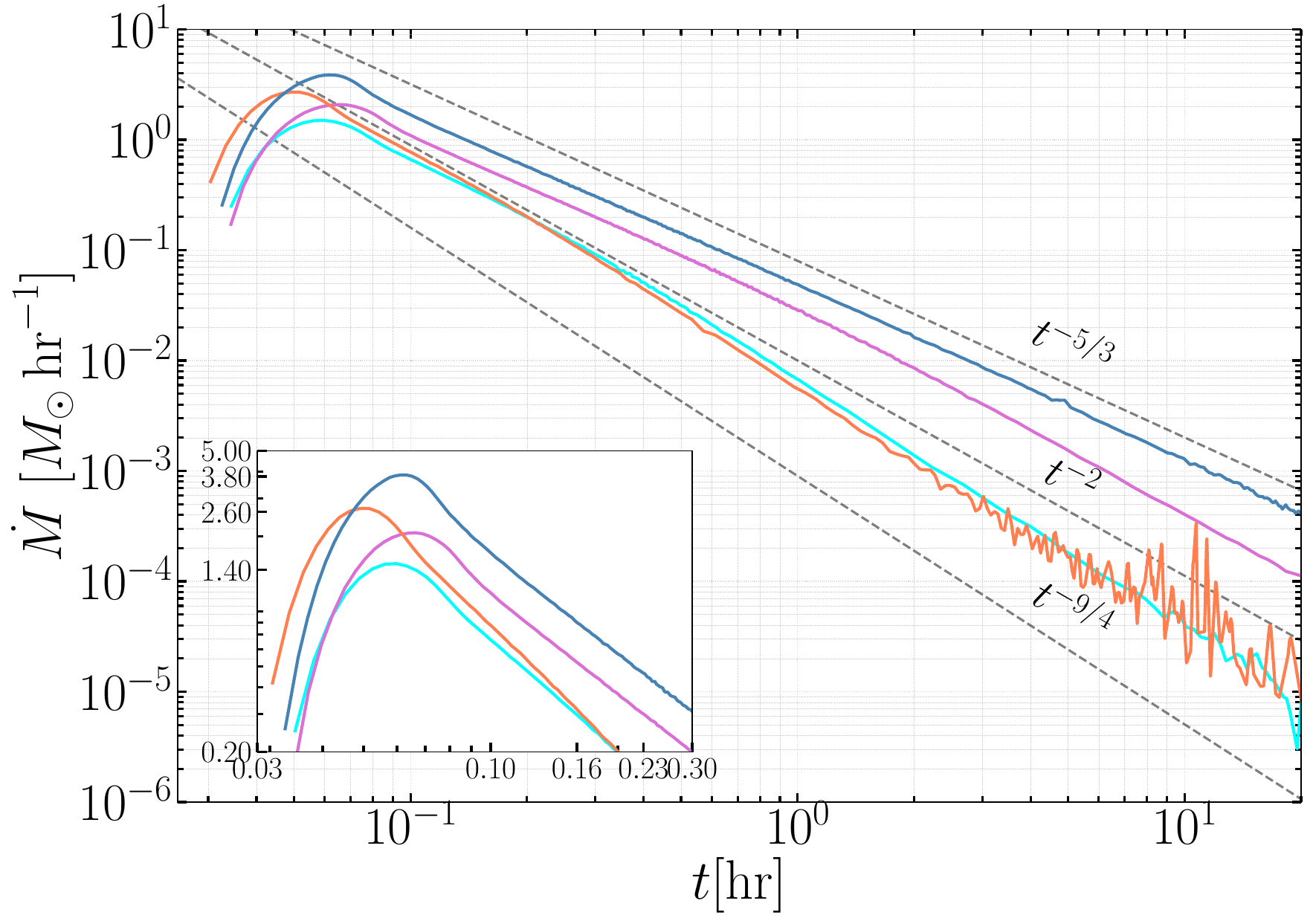}
	\end{subfigure}\\
	\vspace{0.2cm}
	\includegraphics[scale=0.9]{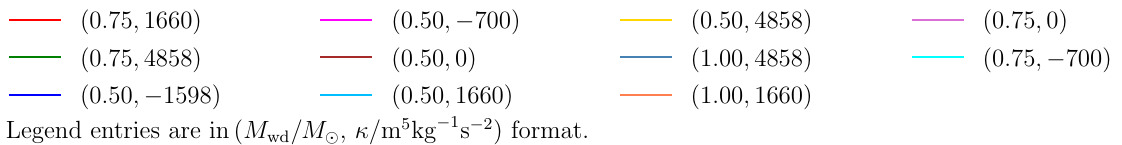}
	\caption{\textbf{Left panel:} The peak fallback rates (in Solar masses per hour) are plotted on a log-log scale as a function of time in hours. The plot includes simulations of different WDs with varying masses and $\kappa$ values for full disruption. At late times, all simulations exhibit a temporal scaling of $t^{-5/3}$ as expected. \textbf{Right panel:} The peak fallback rates (in Solar masses per hour) are plotted against time in hours for different WDs with varying masses and $\kappa$ values. The simulations consider partial disruption scenarios.}
	\label{fig.fallback_obs}
\end{figure}
Another observable related to partial disruption is the trajectory deviation of the core from its initial parabolic trajectory. These deviations arise from the increase in specific orbital energy and specific orbital angular momentum of the core. The Bottom Right panel of Figure \ref{fig.core_obs} displays these deviations, which are directly influenced by the asymmetry. In the zoomed portion, it becomes apparent that the WD with an initial mass of $1.00 M_{\odot}$ and $\kappa = 4858 \, \text{m}^5\text{kg}^{-1}\text{s}^{-2}$ exhibits the highest asymmetry, resulting in the most significant deviation in its core trajectory. The mass ratio, $q$, plays a significant role in influencing the trajectories, as demonstrated in the quantitative analysis by \cite{Banerjee2023}. Our choice of parameters falls with the range found in \cite{Banerjee2023}, thereby producing significant deviations in the trajectory. In the figure, the $x$ and $y$ axes are normalized to the pericenter distance ($r_p = 70 r_g$).

Figure \ref{fig.fallback_obs} illustrates the fallback rates onto the black hole in Solar mass per hour as a function of time in hours. The left panel displays the fallback curves for fully disrupted WDs, while the right panel shows the curves for four partially disrupted WDs. As the initial mass of the WD increases, the magnitude of the peak also increases, indicating a larger amount of debris falling back onto the black hole. For fully disrupted WDs, the late-time slope follows an expected scaling of $t^{-5/3}$ (see \cite{Rees1988}). However, we observe variations in the late-time slope for partially disrupted WDs. Specifically, two WDs with $\beta$ values around $0.72$, corresponding to mass $1.00 M_{\odot}$ with $\kappa = 1660 \, \text{m}^5\text{kg}^{-1}\text{s}^{-2}$ and mass $0.75 M_{\odot}$ with $\kappa = -700 \, \text{m}^5\text{kg}^{-1}\text{s}^{-2}$, exhibit a late-time slope scaling of $t^{-9/4}$, that is in agreement with \cite{Coughlin2019}. Another partially disrupted WD with $\beta = 0.81$, the mass of $0.75 M_{\odot}$ and $\kappa = 0$, initially follows a $t^{-5/3}$ scaling, transitioning to a $t^{-2}$ temporal scaling, and faintly showing a $t^{-9/4}$ behavior at very late times. Finally, the WD with $\beta = 0.876$, mass of $1.00 M_{\odot}$, and $\kappa = 4858 \, \text{m}^5\text{kg}^{-1}\text{s}^{-2}$, exhibits a late-time slope of $t^{-5/3}$. In order to gain a comprehensive understanding of the behavior of peak fallback rates at late times, further investigations are required, exploring different parameter regimes that yield varying values of $\beta$. However, we leave this as a topic for future study.


\subsection{Fixed impact parameter ($\beta$) simulations}
\label{fixedbeta}

In this subsection, we explore the results of simulations with fixed impact parameters set at values of $0.80$ and $1.00$ for the WD with a mass of $0.75 M_{\odot}$. These selected values of $\beta$ ensure that both partial and full disruptions occur during the simulations. We remind the reader of the following caveat before we begin the analysis. 
Note that as we have mentioned after Equation (\ref{tidal_eq}) (see the discussion after this equation), the tidal radius computed from this equation is an estimation. 
It does not take into account the stellar structure and hydrodynamics, and should be modified in the presence of EiBI gravity. Hence $\beta = r_t/r_p$ computed from 
Equation (\ref{tidal_eq}) is only approximate and might vary from its presumed fixed value. Nonetheless, within this approximation, fixed $\beta$ simulations give us useful insights as we 
can compare the behaviour of the stars with experiencing approximately the same relative strength of the tidal field at the pericenter. We will proceed with this caveat in mind.

\begin{figure}[H]
	\centering 
	\begin{subfigure}{0.45\textwidth}
		\centering
		\includegraphics[width=\textwidth]{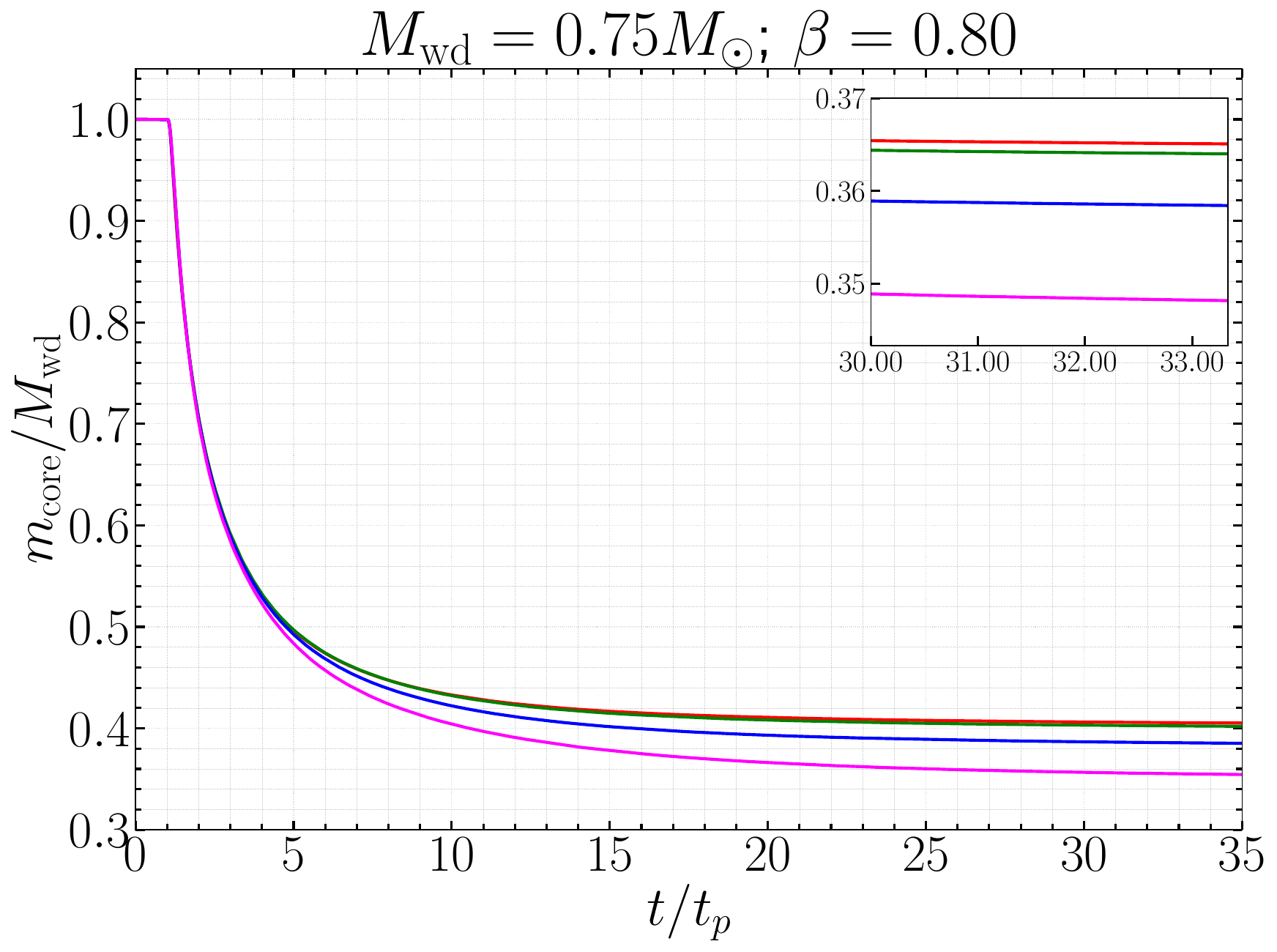}
	\end{subfigure}\hfill
	\begin{subfigure}{0.45\textwidth}
		\centering
		\includegraphics[width=\textwidth]{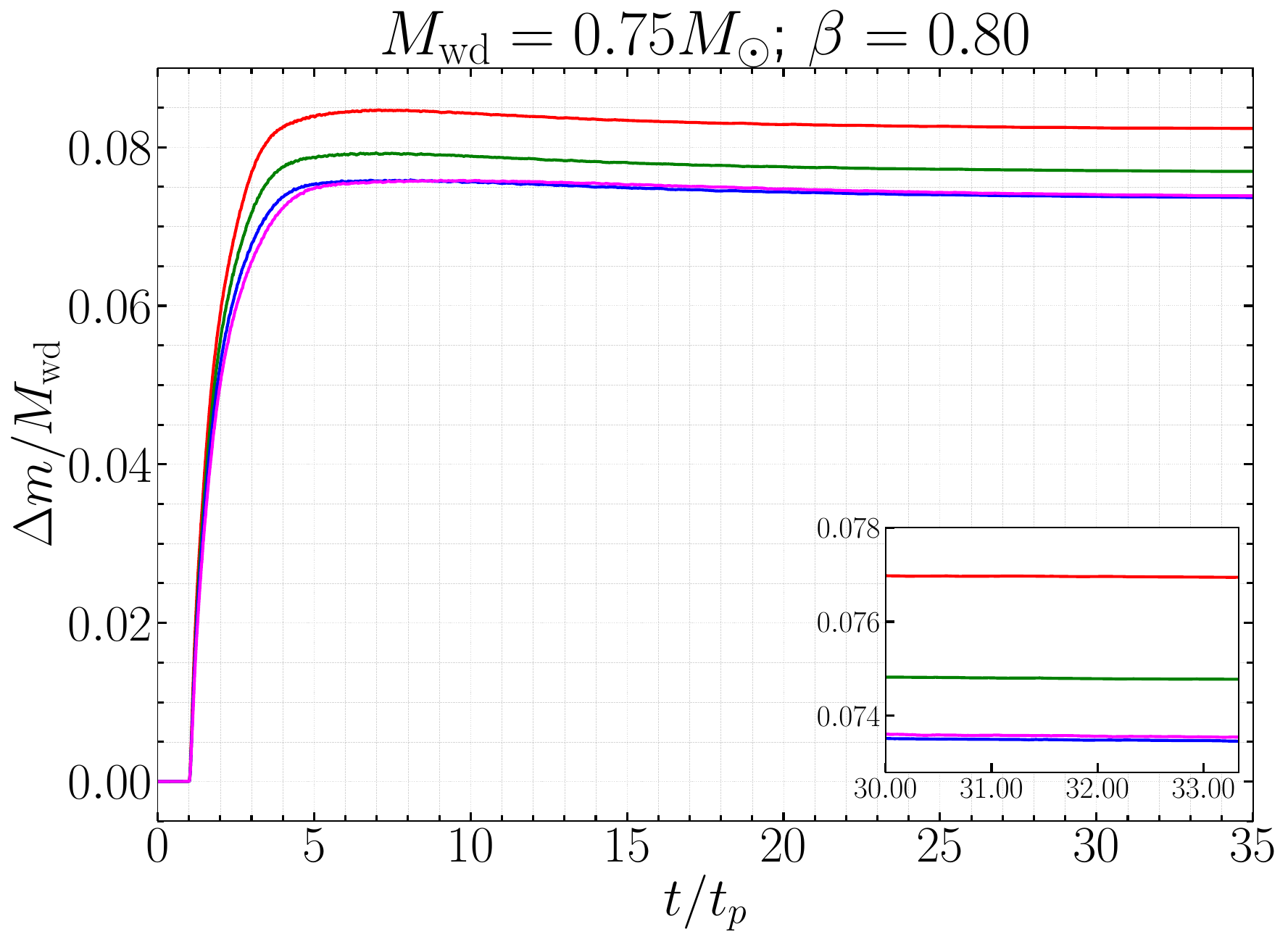}
	\end{subfigure}\\
	\begin{subfigure}{0.45\textwidth}
		\centering
		\includegraphics[width=\textwidth]{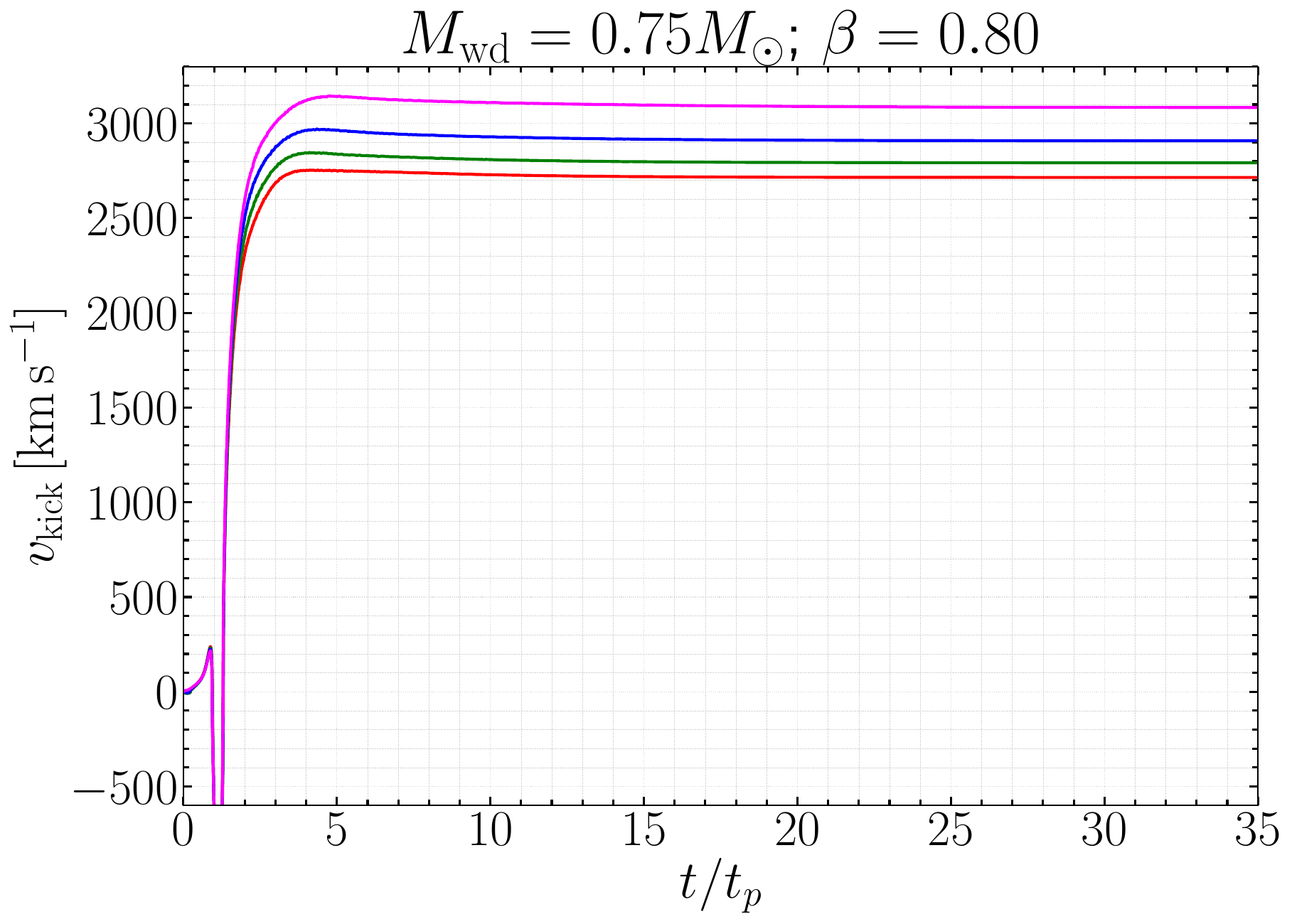}
	\end{subfigure}\hfill
	\begin{subfigure}{0.45\textwidth}
		\centering
		\includegraphics[width=\textwidth]{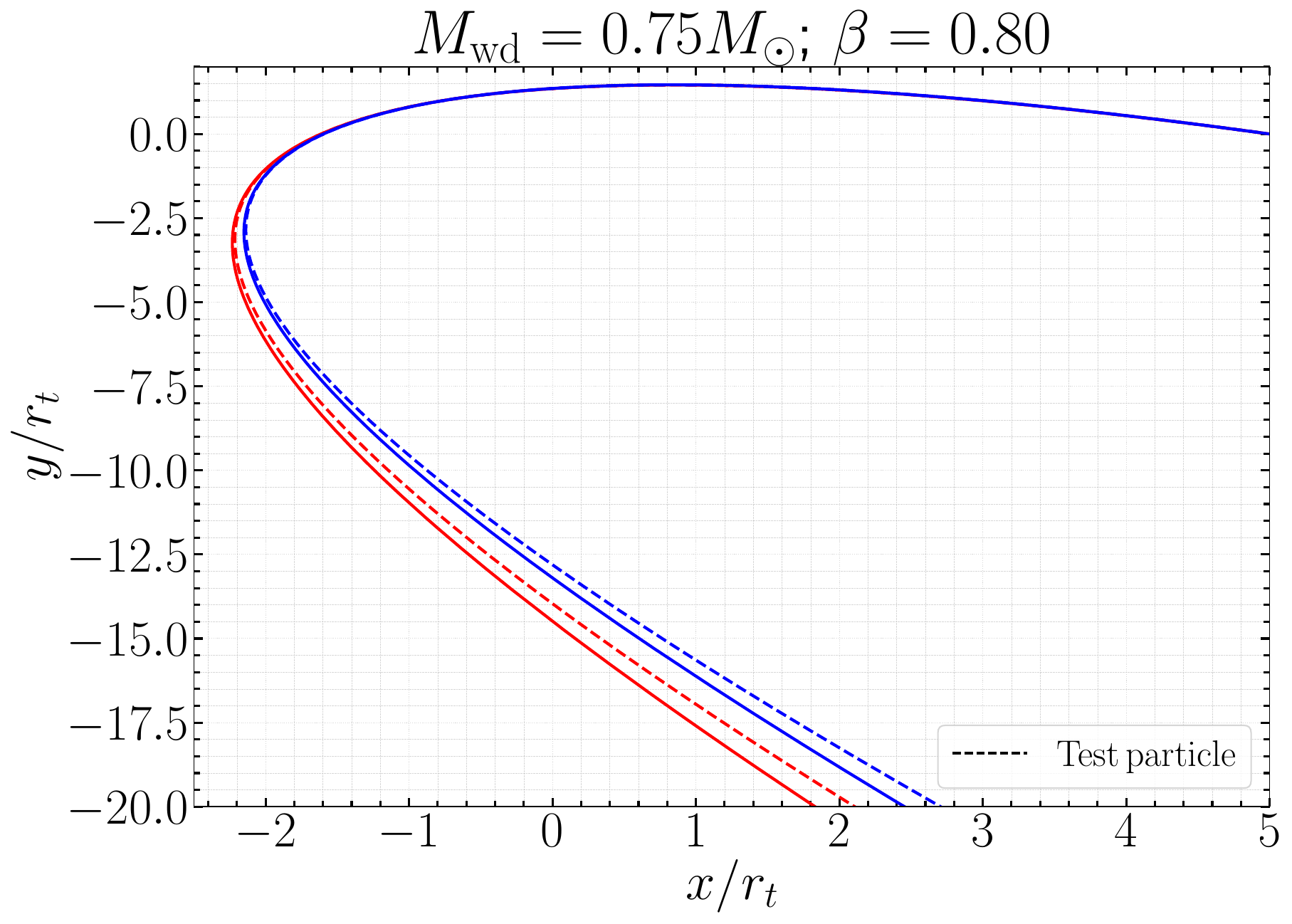}
	\end{subfigure}\\
	\begin{subfigure}{0.45\textwidth}
		\centering
		\includegraphics[width=\textwidth]{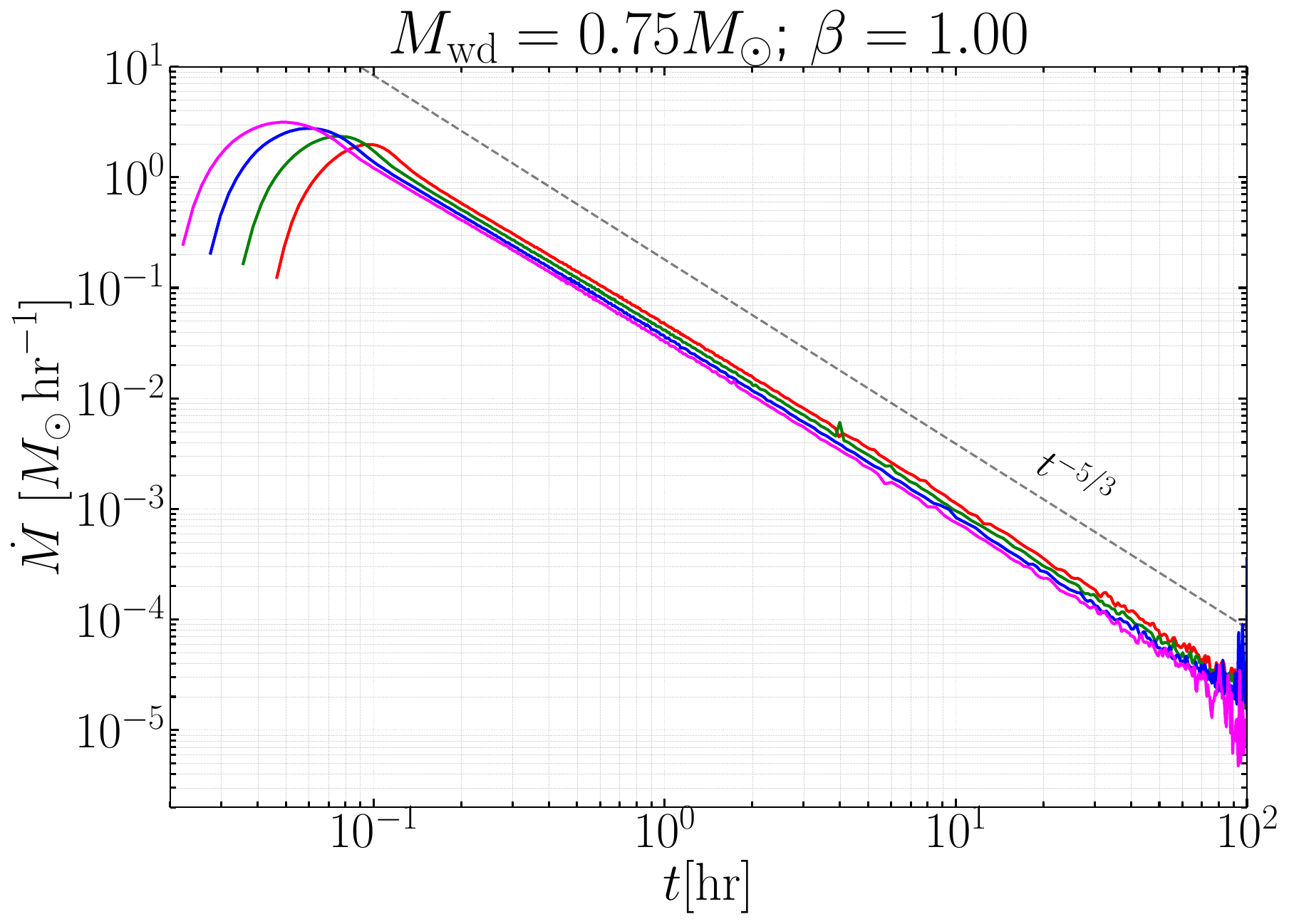}
	\end{subfigure}\hfill
	\begin{subfigure}{0.45\textwidth}
		\centering
		\includegraphics[width=\textwidth]{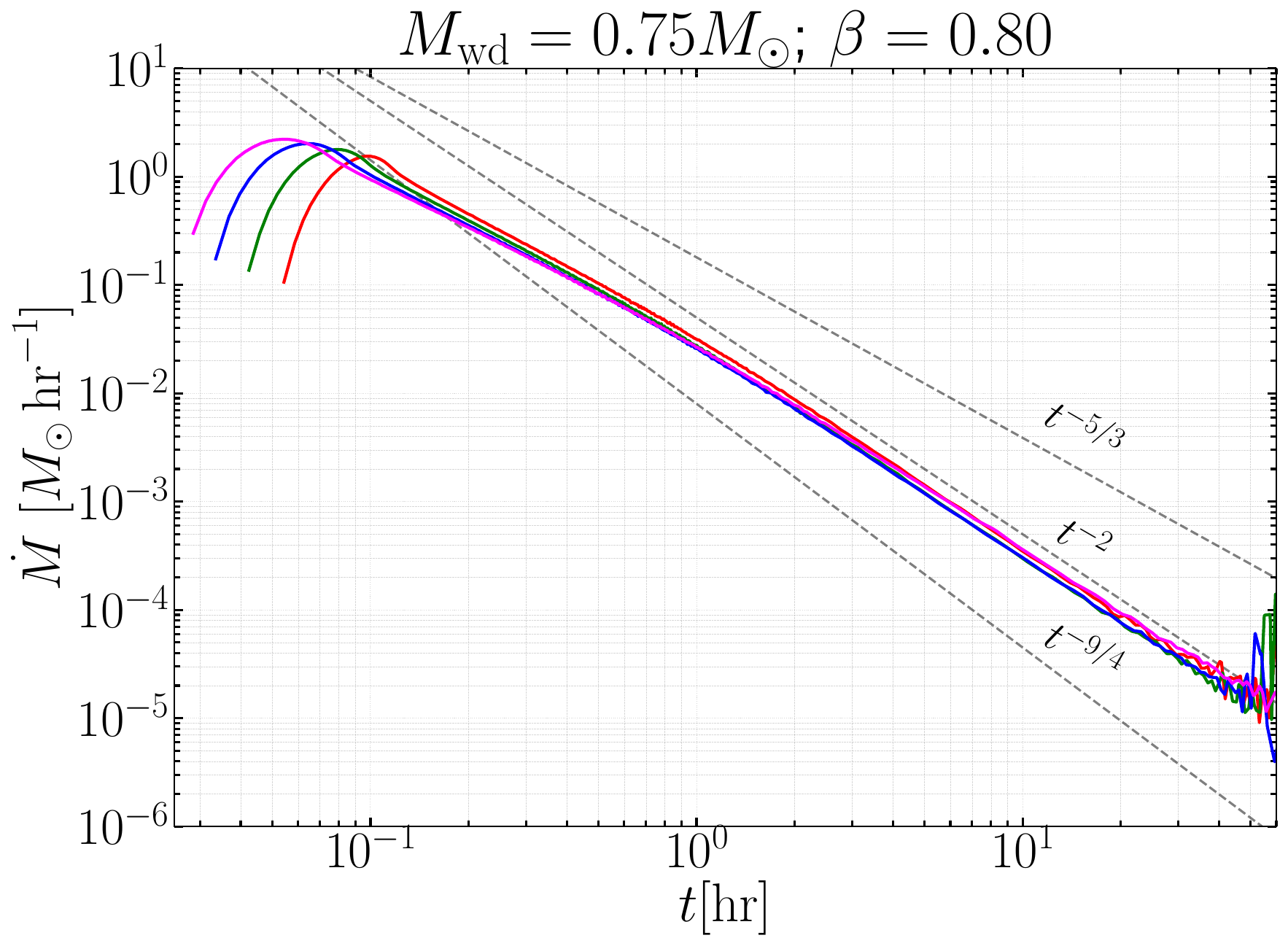}
	\end{subfigure}
	\vspace{0.1cm}
	\includegraphics[scale=0.885]{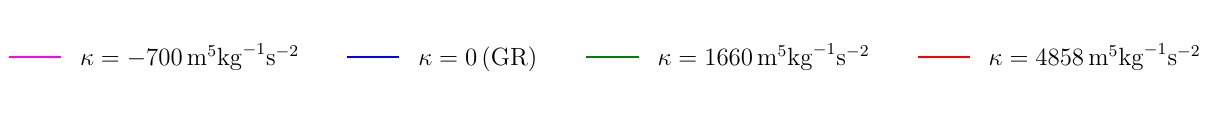}
	\caption{\textbf{Top Left:} Evolution of the bound core mass fraction with normalized time is presented for various WDs mentioned in the legend. \textbf{Top Right:} Variation of the mass difference between two tidal tails with time is shown for various WDs mentioned in the legend. \textbf{Middle Left:} Plot of the kick velocity of the core over time. \textbf{Middle Right:} Trajectory deviations for two stars from the initial parabolic trajectory are presented.
	\textbf{Bottom Left \& Right:} The peak fallback rates are plotted against time for different $\kappa$ values with $\beta = 1.00, 0.80$ respectively.}
	\label{fig.core}
\end{figure}

Starting with the Top Left panel of Figure \ref{fig.core}, we present the time evolution of the bound core mass fraction. Significantly, an increase in the modified gravity parameter $\kappa$ results in an increased bound core mass fraction. 
Additionally, we observe a deviation in the core mass corresponding to different $\kappa$ values from the core mass at $\kappa = 0$. For instance, the deviation increases up to $4.8\%$ when $\kappa$ rises from $\kappa = 0$ to $\kappa = 4858 \,\text{m}^5\text{kg}^{-1}\text{s}^{-2}$. Furthermore, we study the mass difference between the two tidal tails in the Top Right panel of Figure \ref{fig.core}. Interestingly, we find that as $\kappa$ increases, the mass difference between the tails also increases. 
We also analyze the variation of kick velocity over normalized time in the Middle Left panel of Figure \ref{fig.den}. Notably, the kick velocity exhibits an opposite trend compared to the mass difference between the tails. This is due to the fact that after the tidal interaction, the core orbital energy and mass both increases with higher $\kappa$. However, the gain in mass exceeds the gain in orbital energy, resulting in a lower specific orbital energy gain for higher $\kappa$ value. 

Due to the formation of asymmetric tails during partial disruption, there are changes in the core's specific orbital energy and angular momentum, subsequently altering the trajectory of the bound core's center of mass from its initial trajectory. These trajectory deviations are depicted in the Middle Right panel of Figure \ref{fig.den}. While we obtained trajectory deviations for all $\kappa$ values, we chose to present the deviations specifically for $\kappa = 4858 \, \text{m}^5\text{kg}^{-1}\text{s}^{-2}$ and $\kappa = 0$ for better visualization. These specific values 
help us highlight the variations in the trajectory deviations more clearly. Notably, the $x$ and $y$ axes in the plots are normalized to the tidal radius of the respective stars. The figure clearly demonstrates that higher $\kappa$ values correspond to greater trajectory deviations. This is because with increasing $\kappa$, however, the specific orbital energy is lower, but there is a prominent increase in specific angular momentum. The higher change in specific angular momentum deviates the trajectory more for a higher $\kappa$ value.

In the Bottom panels of Figure \ref{fig.core}, we present the behaviors of the fallback rates in both partial (Right panel) and full disruption (Left panel) scenarios, considering varying values of $\kappa$. It is observed that increasing $\kappa$ leads to a decrease in the peak magnitude of the fallback rate, along with an increase in the time of peak and the return time of the most bound debris. The observed trend can be explained by the increase in $r_p$ as $\kappa$ rises to maintain a fixed $\beta$. As a result, the tidal field strength acting on the less compact star, which has a higher $\kappa$, diminishes, resulting in a decreased amount of material being torn apart from the star. Additionally, in full simulations, the temporal scaling at late times follows a $t^{-5/3}$ power law. However, in partial disruption scenarios, the presence of the core introduces a deviation from the $t^{-5/3}$ slope. After the peak, all partial disruption simulations initially exhibit a $t^{-5/3}$ scaling for a few hours, but the decline subsequently steepens and transitions to a $t^{-2}$ slope, eventually reaching a $t^{-9/4}$ decline at the end.

\section{Discussion and Summary}
\label{sec5}

The methods of smoothed particle hydrodynamics provide an invaluable tool to study stellar dynamics and has been immensely popular over the decades, and has
provided several useful insights therein. In this work we have extended the scope of SPH further, by incorporating the effects of a class of modified gravity theories --
in particular to study tidal disruption dynamics of WDs in the background of intermediate-mass black holes. The interior of the WDs have been  
modelled by incorporating EiBI gravity in this study. Where there are several works in the literature that seek to constrain modified gravity using astrophysical
tests, here we have used an allowed range of parameters and explored the effects of modified gravity in a realistic tidal disruption scenario, and we have
quantified how various tidal disruption events observables depends on modified gravity. We believe that this is the first work of this kind to appear in the
literature.  As we have mentioned in the introduction, any modification of gravity is associated with possible extra degrees of freedom and leaves a low energy 
imprint via parameters ($\kappa$ in our case) that typically alter the pressure balance equation inside stellar objects. In this sense,  our work can be thought of as generic, and should be applicable to a wide range of modified gravity theories, the caveat being that assumptions of spherical symmetry might make other theories more challenging than the present study.  

In this paper, we have used a zero temperature EOS to relate the pressure and density, an improvement from the polytropic EOS. As a check, we modelled the 
lower mass WDs ($0.5 M_{\odot}$) using polytropic EOS with $n = 5/3$, where $n$ is the polytropic index, and found that the polytropic EOS gave almost the same results for tidal observables, but as we go towards the higher mass WDs ($1 M_{\odot}$) neither $n=5/3$ nor $n = 1.35$ gave satisfactory results. Thus, zero temperature EOS can be valuable to model higher-mass WDs without assuming a polytropic approximation. 
In this context, note also that we selected WDs with three different masses: $0.50, 0.75, 1.00 M_{\odot}$. The choice of increasing mass values was motivated by the mass-radius relation, which indicates that the deviations from $\kappa = 0$ become more compared to the $5\sigma$ bounds as the mass increases.

In this study, we investigate tidal disruption events involving different white dwarf stars with various $\kappa$ values, employing two different approaches. Firstly, from an observational perspective, we maintain a constant pericenter distance for all white dwarfs, ensuring the same tidal field strength. Interestingly, we find that white dwarfs with different $\kappa$ values display unique behaviors in partial disruptions. With increasing $\kappa$, the initial white dwarfs with the same mass experience greater mass loss, and the mass difference between the two tails become more due to deeper encounters. This asymmetric mass loss induces a kick velocity to the remnant core, which increases with $\kappa$ and can reach values of up to $\sim 5\times10^3 \, \text{km}\,\text{s}^{-1}$ for an initial white dwarf mass of $1.00 M_{\odot}$. Moreover, the specific energy and specific angular momentum changes of the core lead to deviations from its initial trajectory, and these deviations are also observed to increase with $\kappa$. Additionally, the peak magnitude, time of the peak, and return time of the most bound debris show variations among white dwarfs with different $\kappa$ values.

Furthermore, we conduct simulations with a fixed impact parameter, $\beta$, for white dwarfs of mass $0.75 M_{\odot}$. Although the determination of $\beta$ requires the use of an approximate formula for the tidal radius (Equation (\ref{tidal_eq})), we explore the behavior of observables in both partial and full disruptions within this approximation. In the case of partial disruption, the core mass can vary depending on the $\kappa$ value. For a white dwarf with a mass of $0.75 M_{\odot}$, we find that the core mass increases by approximately $5\%$ when $\kappa$ is increased from $0$ to $4858 \, \text{m}^5\text{kg}^{-1}\text{s}^{-2}$. We also performed additional simulations with a white dwarf of mass $1 M_{\odot}$ and observed an increase in the core mass of up to $20\%$ for the same $\kappa$ values, which represents a significant change allowed by this class of modified gravity theories. Additionally, as $\kappa$ increases, the asymmetry in the two tidal tails induces a kick velocity in the core, resulting in deviations of its trajectory from the initial trajectory. Regarding the peak fallback rate, we find that as $\kappa$ increases, the peak magnitude decreases while the time of peak and return time of the most bound debris increase. The difference in peak magnitude between $\kappa = 4858 \, \text{m}^5\text{kg}^{-1}\text{s}^{-2}$ and $\kappa = 0$ reaches up to $28.5\%$ for $\beta = 1.0$ and up to $23.5\%$ for $\beta = 0.80$ for the $0.75 M_{\odot}$ white dwarf.

These results demonstrate the impact of EiBI gravity on the observables in tidal disruption events. In the near future, upcoming missions like LISA (Laser Interferometer Space Antenna) will provide valuable observational data, offering an opportunity to compare various numerical models for tidal disruptions. In this context, our analysis holds significance as it allows us to explore different effects of modified gravity, such as those predicted by the EiBI theory. By studying the influence on the dynamics and observables of tidal disruption events with different modified gravity parameters, our analysis contributes to a better understanding of gravitational theories beyond the standard framework.

As always, it is useful to analyse possible degeneracies that can arise in our analysis, from other effects. Here, we have taken a well known EOS of WDs,
so that changes to the EOS (compared to say polytropic ones) are not relevant. Further, in the mass range that we consider, magnetic fields are not
known to play a significant role. The only other physical variable that we need to analyse is stellar rotation. SPH in the presence of such rotation
was recently analysed in \cite{Golightly2}, where it was found that the direction of stellar spin helps (hinders) tidal disruption depending on whether
the spin is prograde (retrograde). Note that as we have mentioned in the introduction, the effect of $\kappa$ is {\it qualitatively} similar, i.e., it either strengthens
or reduces gravity depending on its sign, see Equation (\ref{hydro_eq}). Crucially however, the tidal radius is also non-trivially 
modified by $\kappa$, see Equation (\ref{tidal_eq}). To simplify the analysis, let us consider the situation for a fixed impact parameter $\beta$, with a positive $\kappa$. 
Then although a star of a certain mass is less compact compared to the Newtonian case, as EiBI gravity causes an increase in its radius compared to Newtonian 
values, it is also being disrupted at a greater distance,
as $r_t$ and hence $r_p$ both increase due to the effect of EiBI gravity to keep $\beta$ fixed. These two factors together results in the later occurence of the peak fall back
rate with a diminishing magnitude, as is apparent from Figure \ref{fig.core}. 
A detailed analysis of the interplay between modified gravity and stellar rotations is an issue that is worth investigating in the future. \\

\noindent
\begin{center}
{\bf Acknowledgements}\\
\end{center}
\noindent
We acknowledge the support and resources provided by PARAM Sanganak under the National Supercomputing Mission, Government of India, at the Indian Institute of Technology Kanpur. The work of DG is supported by grant number 09/092(1025)/2019-EMR-I from the Council of Scientific and Industrial Research (CSIR). PB acknowledges financial support from Science and Engineering Research Board, Government of India, File Number PDF/2022/000332.  

\bigskip

\noindent
{\bf Data Availability Statement}\\

\noindent
The data underlying this article will be shared upon reasonable request to the corresponding author.

\appendix
\section{Appendix A}
\label{appA}

The M6 kernel function used in SPH, is given by (see \cite{PHANTOM}): 
\begin{align}
	W(r,h) =
	\begin{cases}
		\frac{1}{\pi h^3}\left(\frac{11}{20} - \frac{x^2}{2} + \frac{x^4}{4} - \frac{x^5}{12} \right) & 0 \leq x \leq 1 \\
		\frac{1}{\pi h^3}\left(\frac{17}{40} + \frac{5x}{8} - \frac{7x^2}{4} + \frac{5x^3}{4} - \frac{3x^4}{8} + \frac{x^5}{24} \right) & 1 \leq x \leq 2 \\
		\frac{1}{\pi h^3}\left(\frac{81}{40} - \frac{27x}{8} + \frac{9x^2}{4} - \frac{3x^3}{4} + \frac{x^4}{8} - \frac{x^5}{120} \right) & 2 \leq x \leq 3 \\
		0 & x \geq 3
	\end{cases}
\end{align}
For this M6 kernel, the derivative of the gravitational softening kernel, $\frac{\partial\phi(r,h)}{\partial r}$, is given below:
\begin{align}
	\frac{\partial\phi(r,h)}{\partial r} =
	\begin{cases}
		\frac{1}{h^2}\left(\frac{11x}{15} - \frac{2x^3}{5} + \frac{x^5}{7} - \frac{x^6}{24}\right) \\
		\hspace{1.2cm} + \frac{\kappa'}{\pi h^4}\left(-x + x^3 - \frac{5x^4}{12}\right) &  0 \leq x \leq 1 \\
		\frac{1}{h^2}\left(\frac{1}{336 x^2} + \frac{17 x}{30} + \frac{5 x^2}{8} - \frac{7 x^3}{5} + \frac{5 x^4}{6} - \frac{3 x^5}{14} + \frac{x^6}{48}\right) \\
		\hspace{1.2cm} + \frac{\kappa'}{\pi h^4}\left(\frac{5}{8} - \frac{7 x}{2} + \frac{15 x^2}{4} - \frac{3 x^3}{2} + \frac{5 x^4}{24}\right) &  1 \leq x \leq 2 \\
		\frac{1}{h^2}\left(-\frac{169}{560 x^2} + \frac{27 x}{10} - \frac{27 x^2}{8} + \frac{9 x^3}{5} - \frac{x^4}{2} + \frac{x^5}{14} - \frac{x^6}{240}\right) \\
		\hspace{1.2cm} + \frac{\kappa'}{\pi h^4}\left(-\frac{27}{8} + \frac{9 x}{2} - \frac{9 x^2}{4} + \frac{x^3}{2} - \frac{x^4}{24}\right) &  2 \leq x \leq 3 \\
		\frac{1}{r^2} &  x \geq 3
	\end{cases}
\end{align}
Where $x = r/h$, $\kappa' = \kappa/(4 G)$ and the constants are determined to ensure piecewise continuity of the softening kernel. Finally, the gravitational softening kernel for the M6 kernel is given by:
\begin{align}
	\phi(r,h) =
	\begin{cases}
		\frac{1}{h}\left(-\frac{239}{210} + \frac{11x^2}{30} - \frac{x^4}{10} + \frac{x^6}{42} - \frac{x^7}{168}\right) \\
		\hspace{1.2cm} +\frac{\kappa'}{\pi h^3}\left(\frac{11}{20} - \frac{x^2}{2} + \frac{x^4}{4} - \frac{x^5}{12}\right) &  0 \leq x \leq 1 \\
		\frac{1}{h}\left(-\frac{473}{420} - \frac{1}{336x} + \frac{17x^2}{60} + \frac{5x^3}{24} - \frac{7x^4}{20} + \frac{x^5}{6} - \frac{x^6}{28} + \frac{x^7}{336}\right) \\
		\hspace{1.2cm} +\frac{\kappa'}{\pi h^3}\left(\frac{17}{40} + \frac{5x}{8} - \frac{7x^2}{4} + \frac{5x^3}{4} - \frac{3x^4}{8} + \frac{x^5}{24}\right) &  1 \leq x \leq 2 \\
		\frac{1}{h}\left(-\frac{243}{140} + \frac{169}{560x} + \frac{27x^2}{20} - \frac{9x^3}{8} + \frac{9x^4}{20} - \frac{x^5}{10} + \frac{x^6}{84} - \frac{x^7}{1680} \right) \\
		\hspace{1.2cm} +\frac{\kappa'}{\pi h^3}\left(\frac{81}{40} - \frac{27x}{8} + \frac{9x^2}{4} - \frac{3x^3}{4} + \frac{x^4}{8} - \frac{x^5}{120}\right) &  2 \leq x \leq 3 \\
		-\frac{1}{r} &  x \geq 3
	\end{cases}
\end{align}
\newpage
\section{Appendix B}
\label{appB}

\begin{figure}[h!]
	\centering
	\includegraphics[scale=0.24]{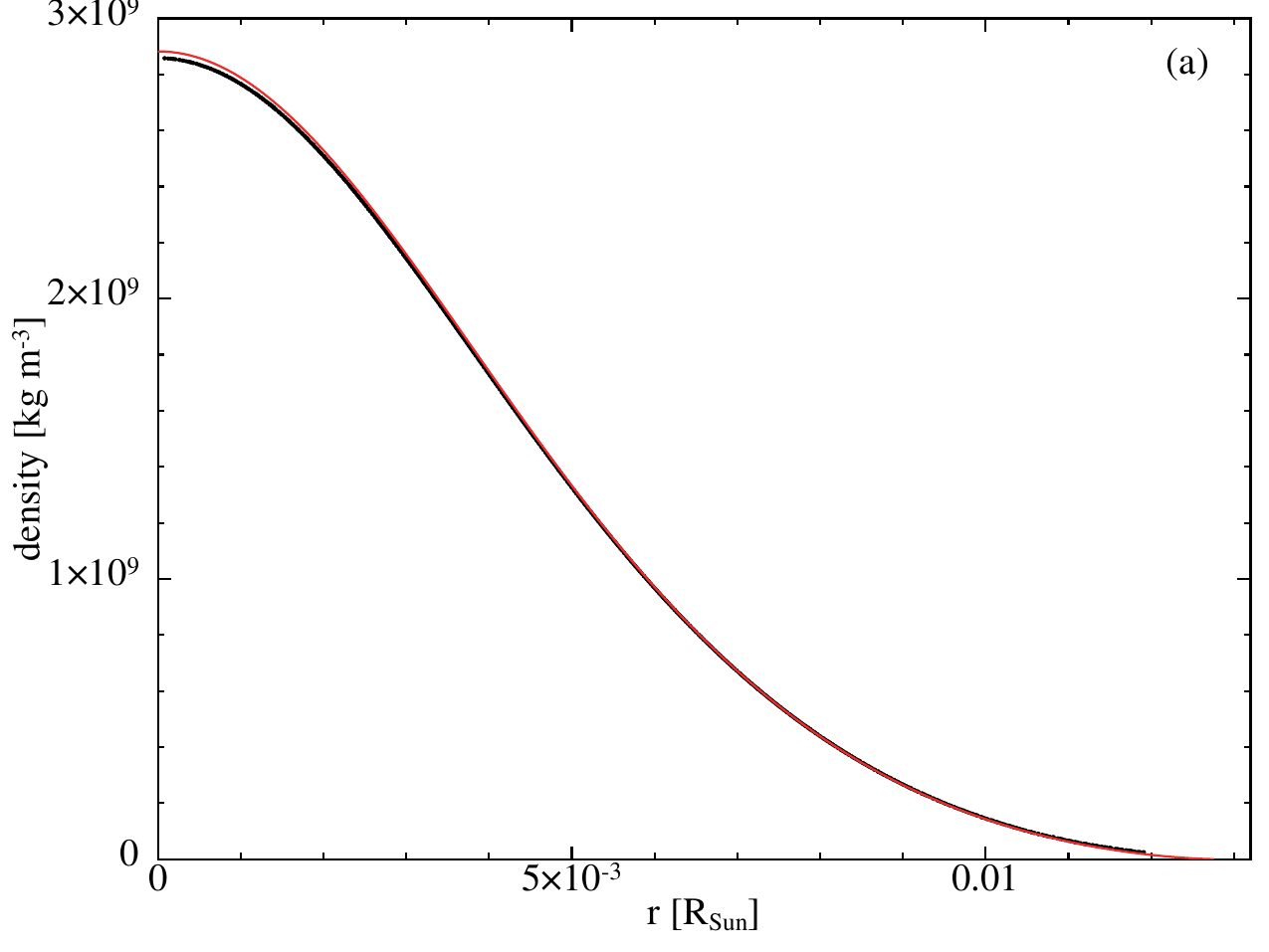}
	\includegraphics[scale=0.24]{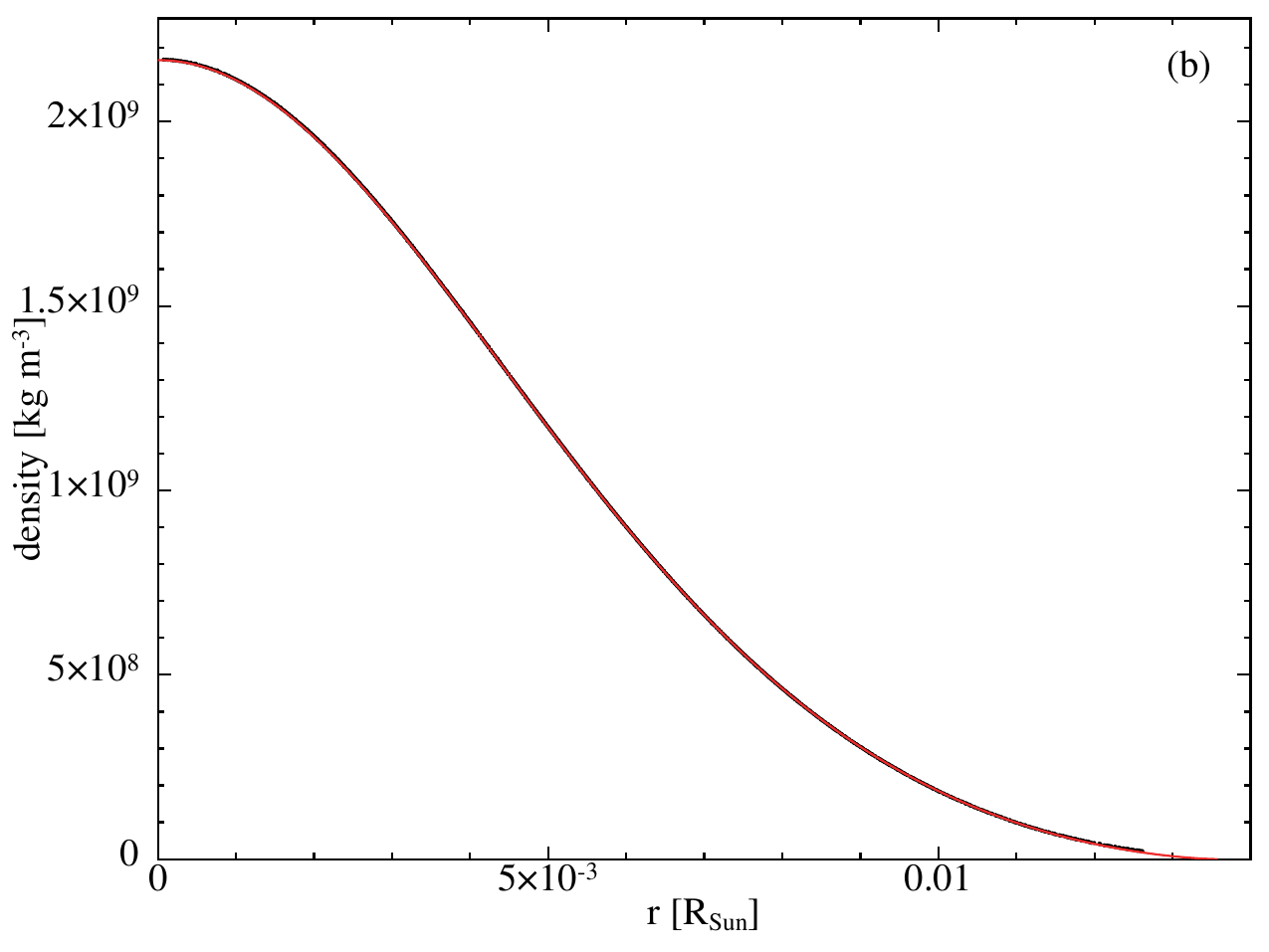}
	\includegraphics[scale=0.24]{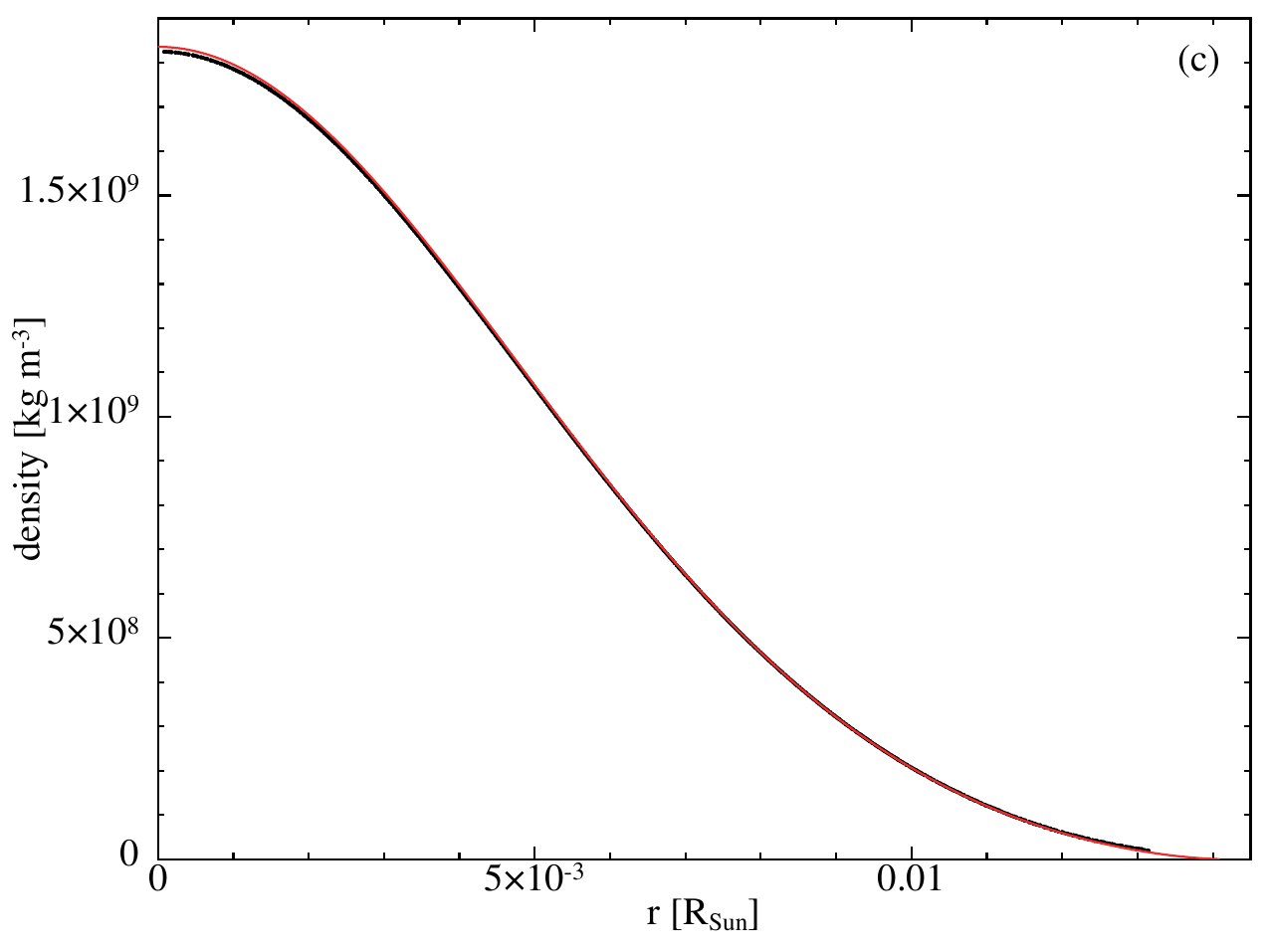}
	\includegraphics[scale=0.24]{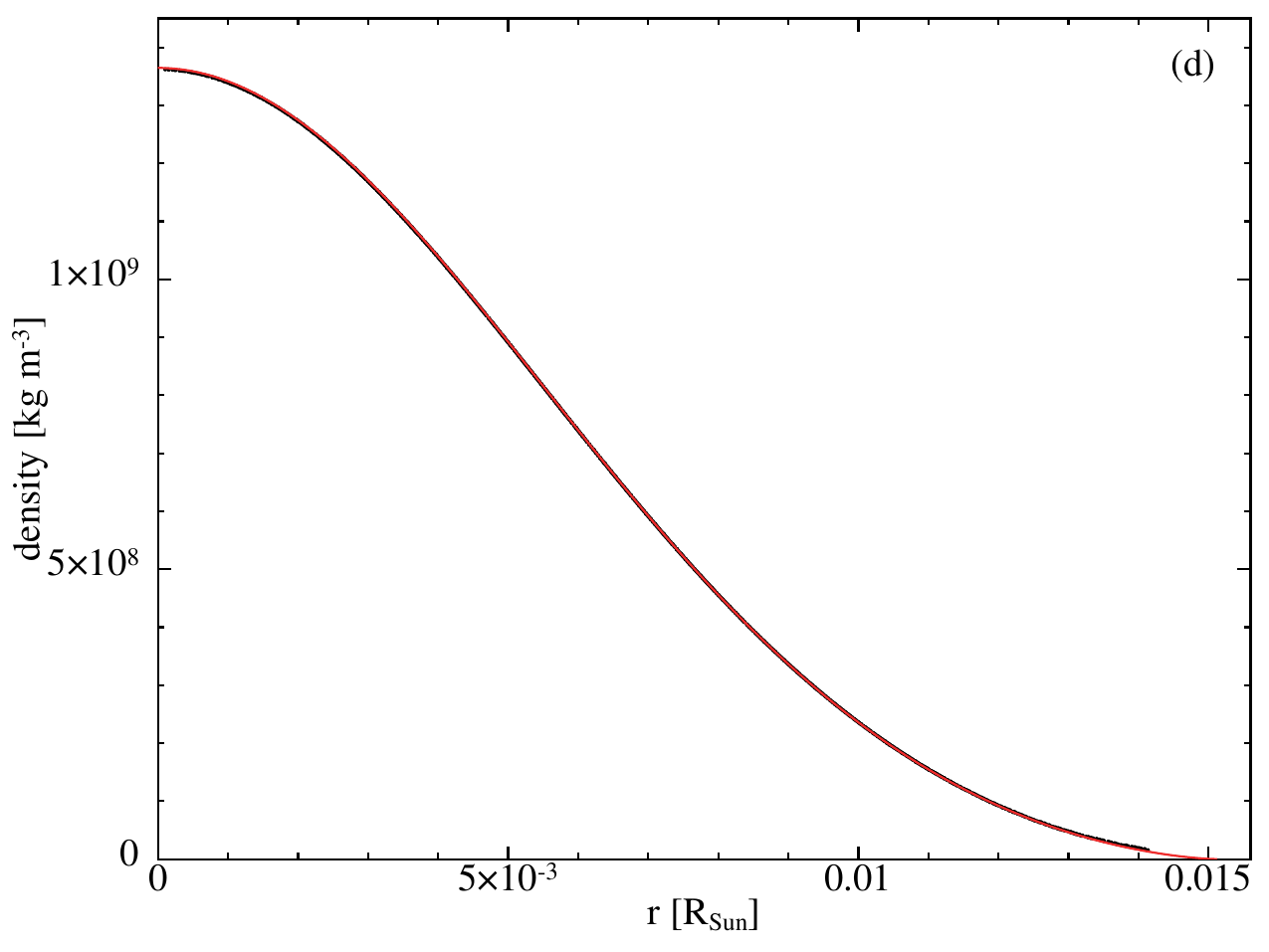}
	\includegraphics[scale=0.24]{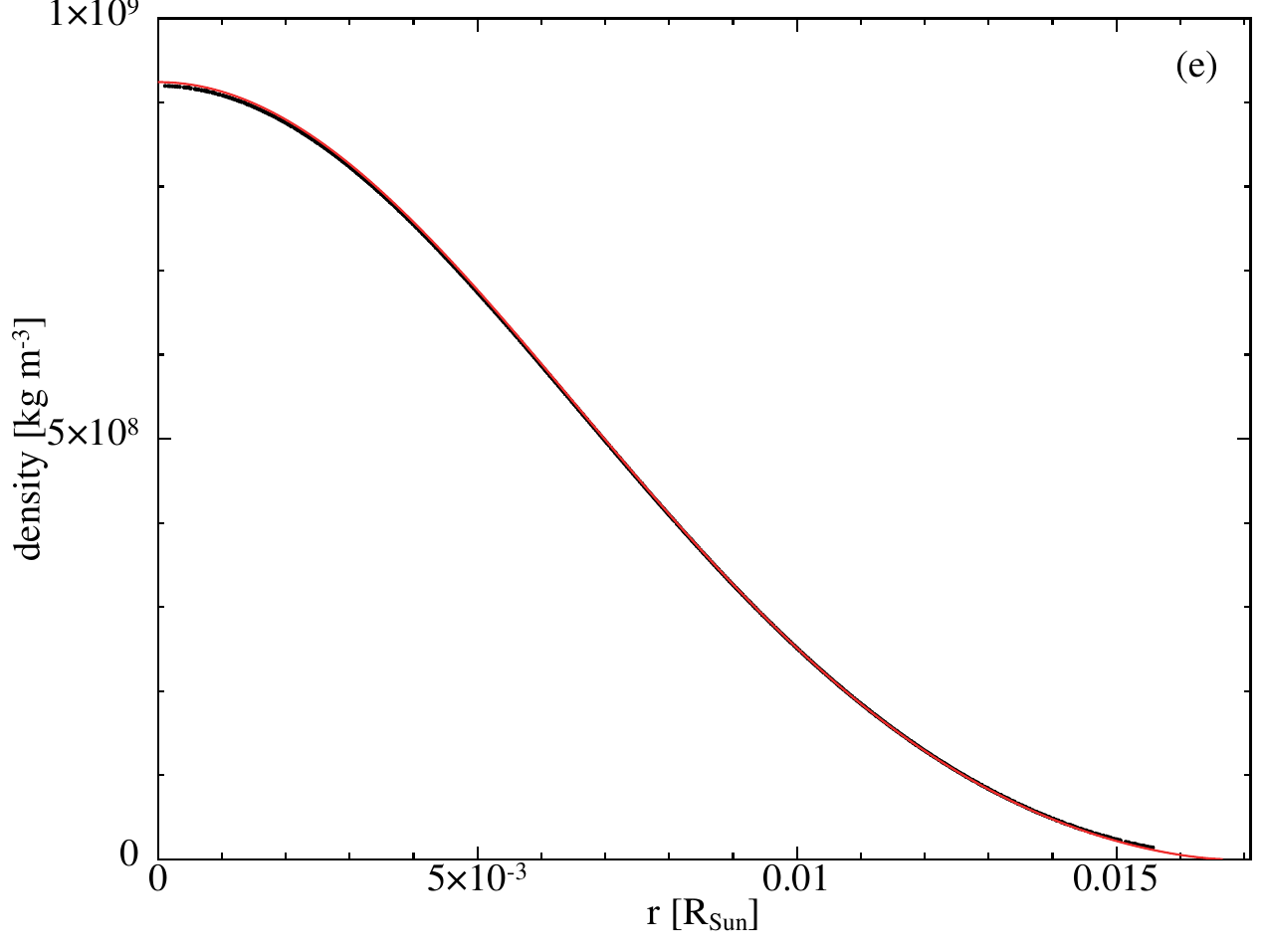}
	\includegraphics[scale=0.24]{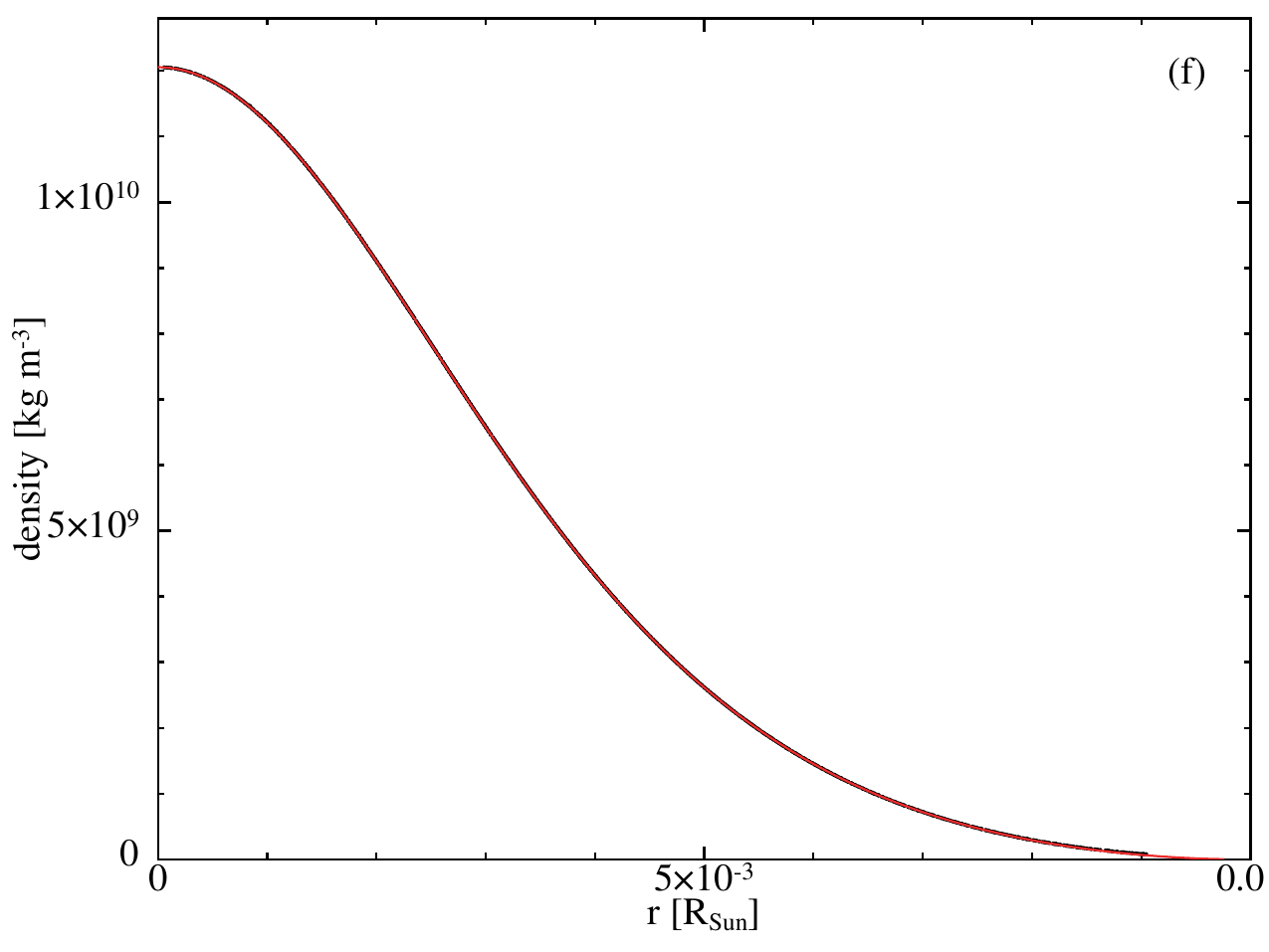}
	\includegraphics[scale=0.24]{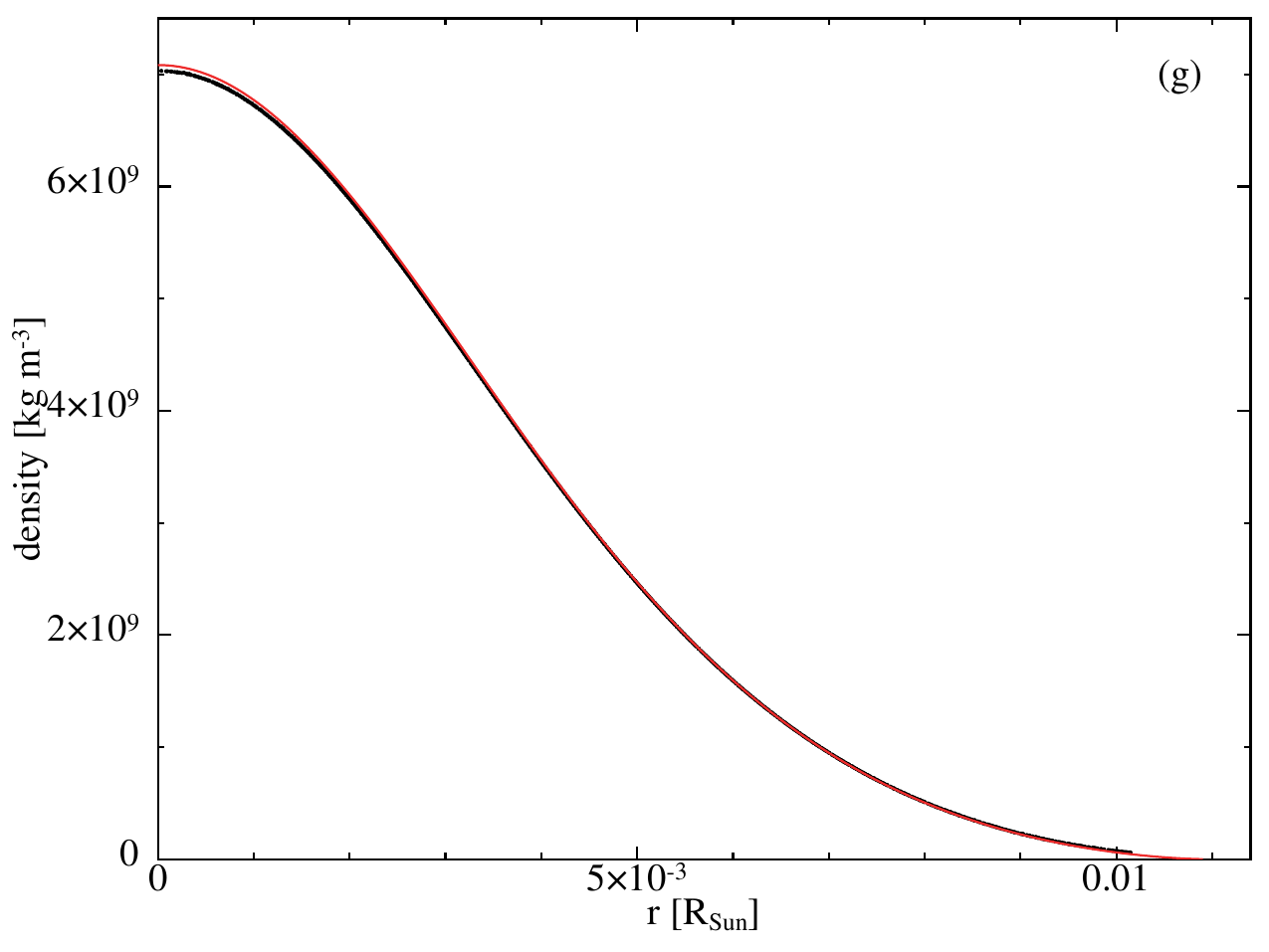}
	\includegraphics[scale=0.24]{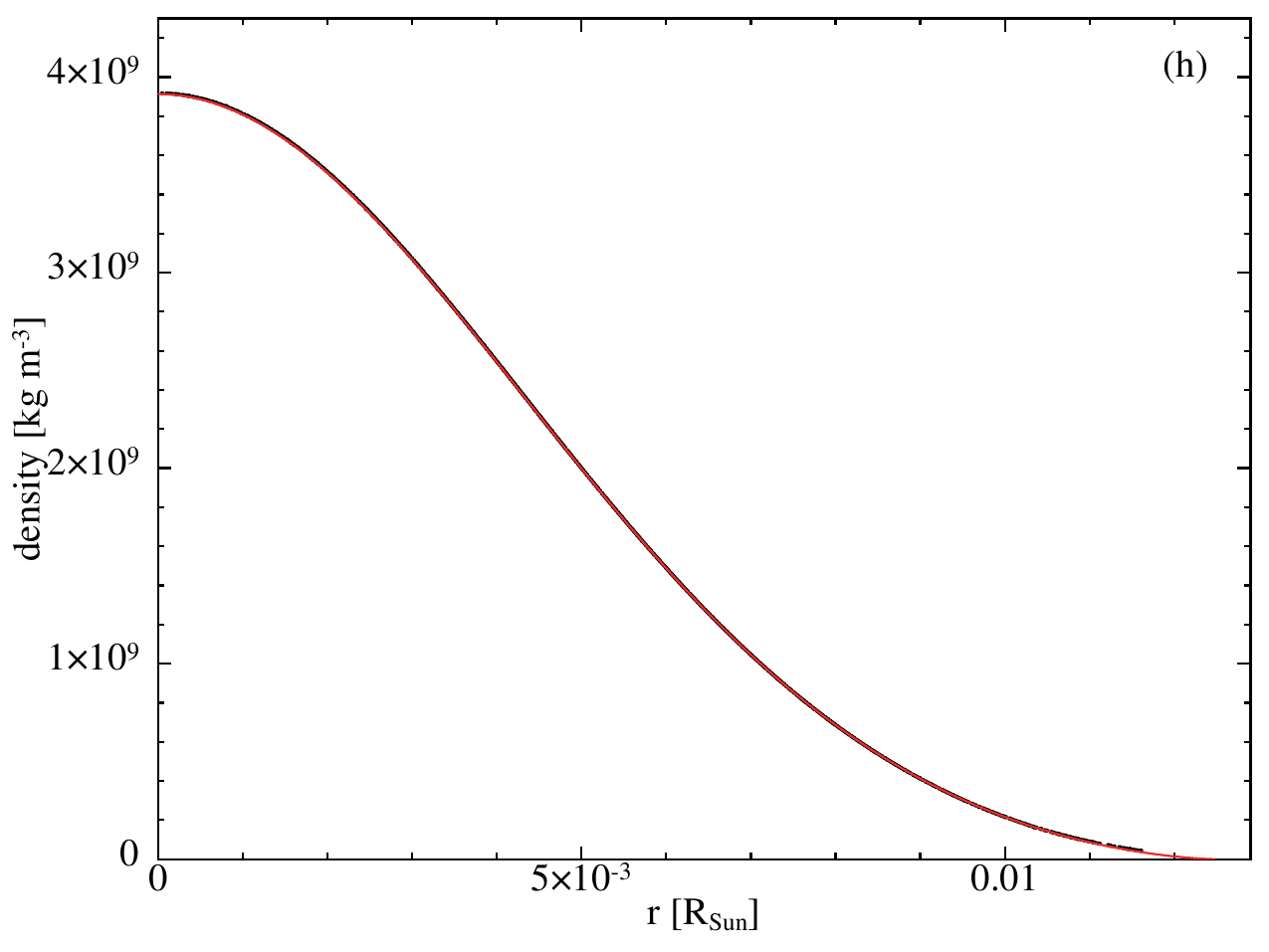}
	\includegraphics[scale=0.24]{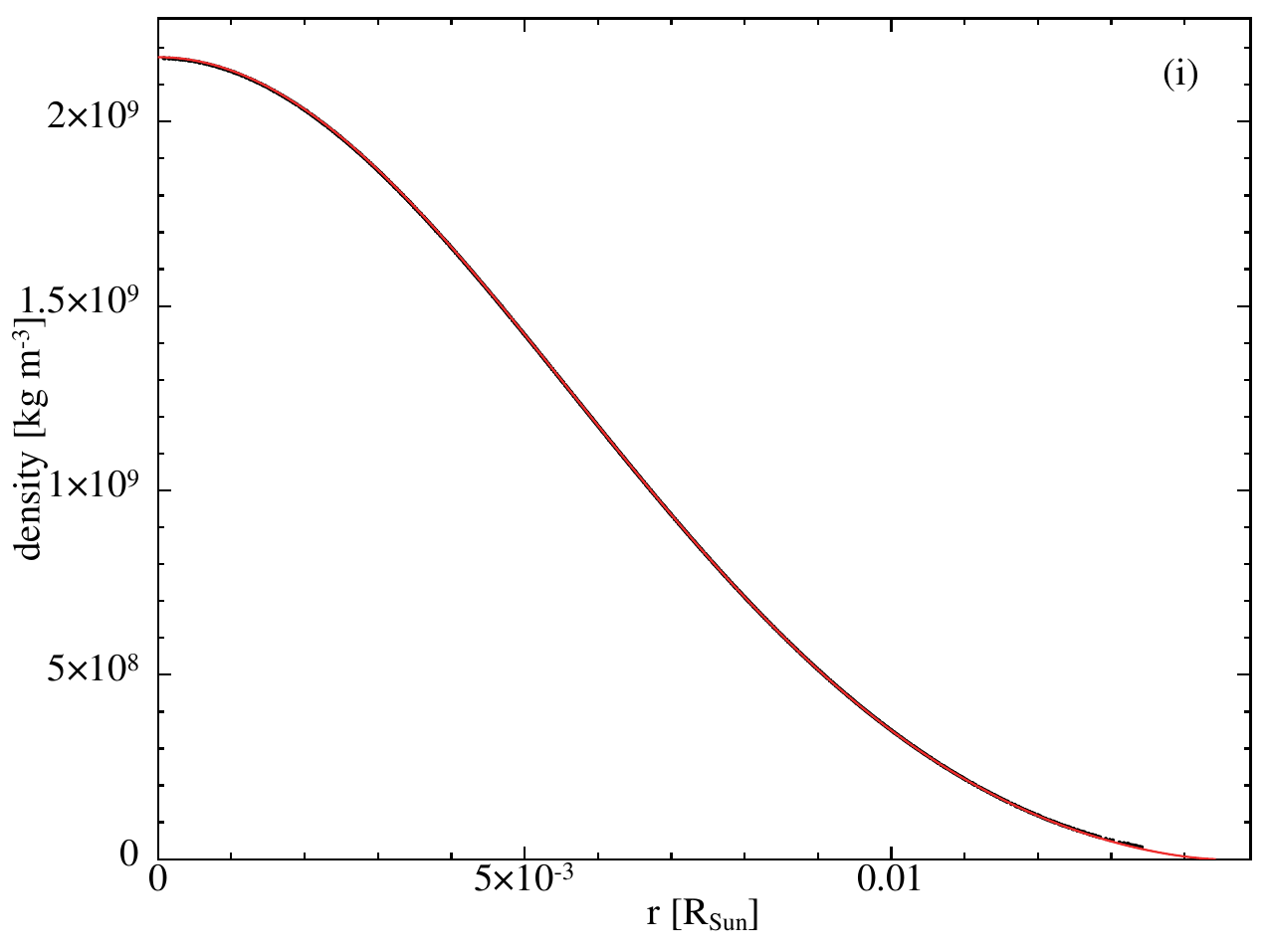}
	\includegraphics[scale=0.24]{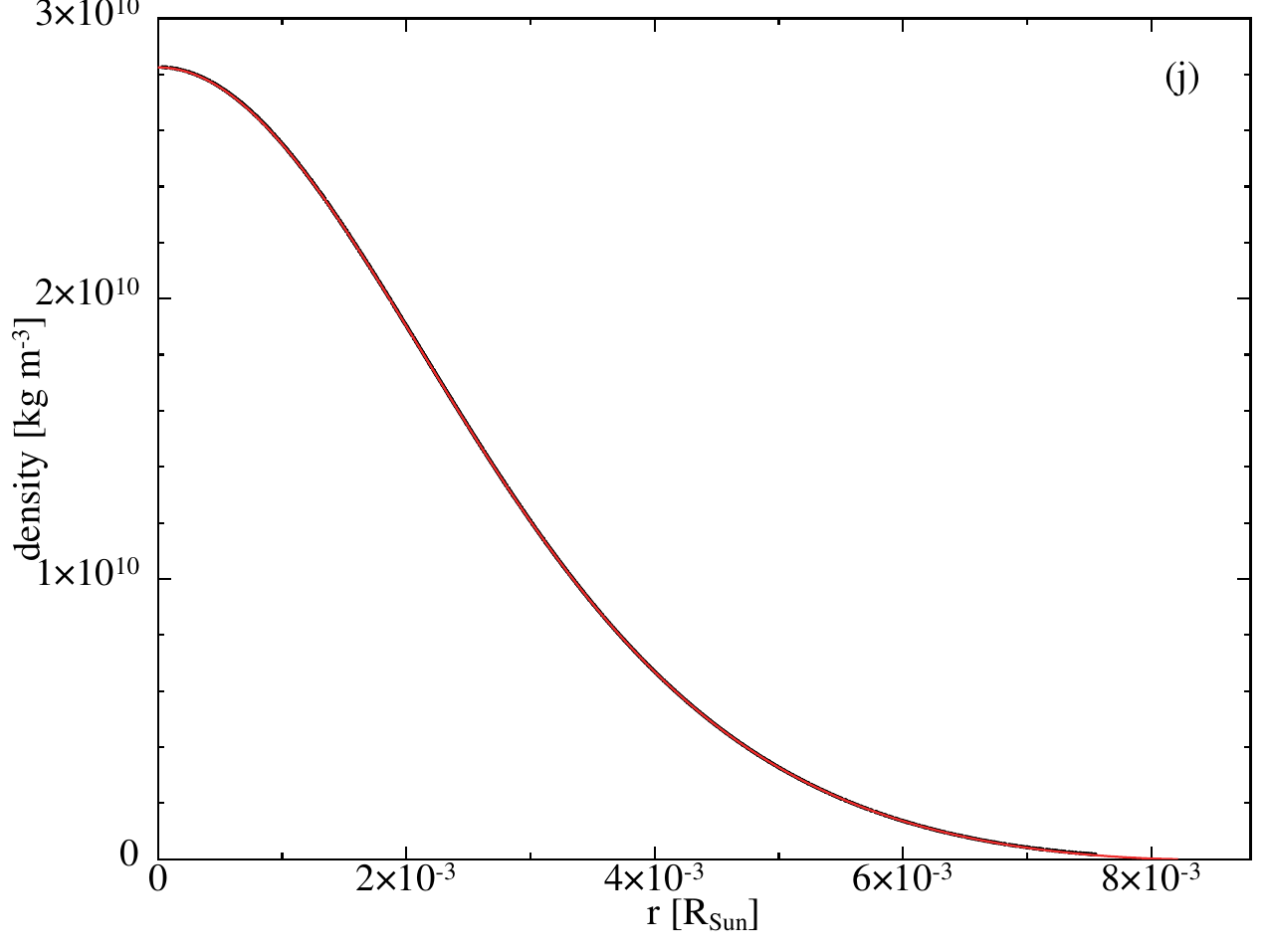}
	\includegraphics[scale=0.24]{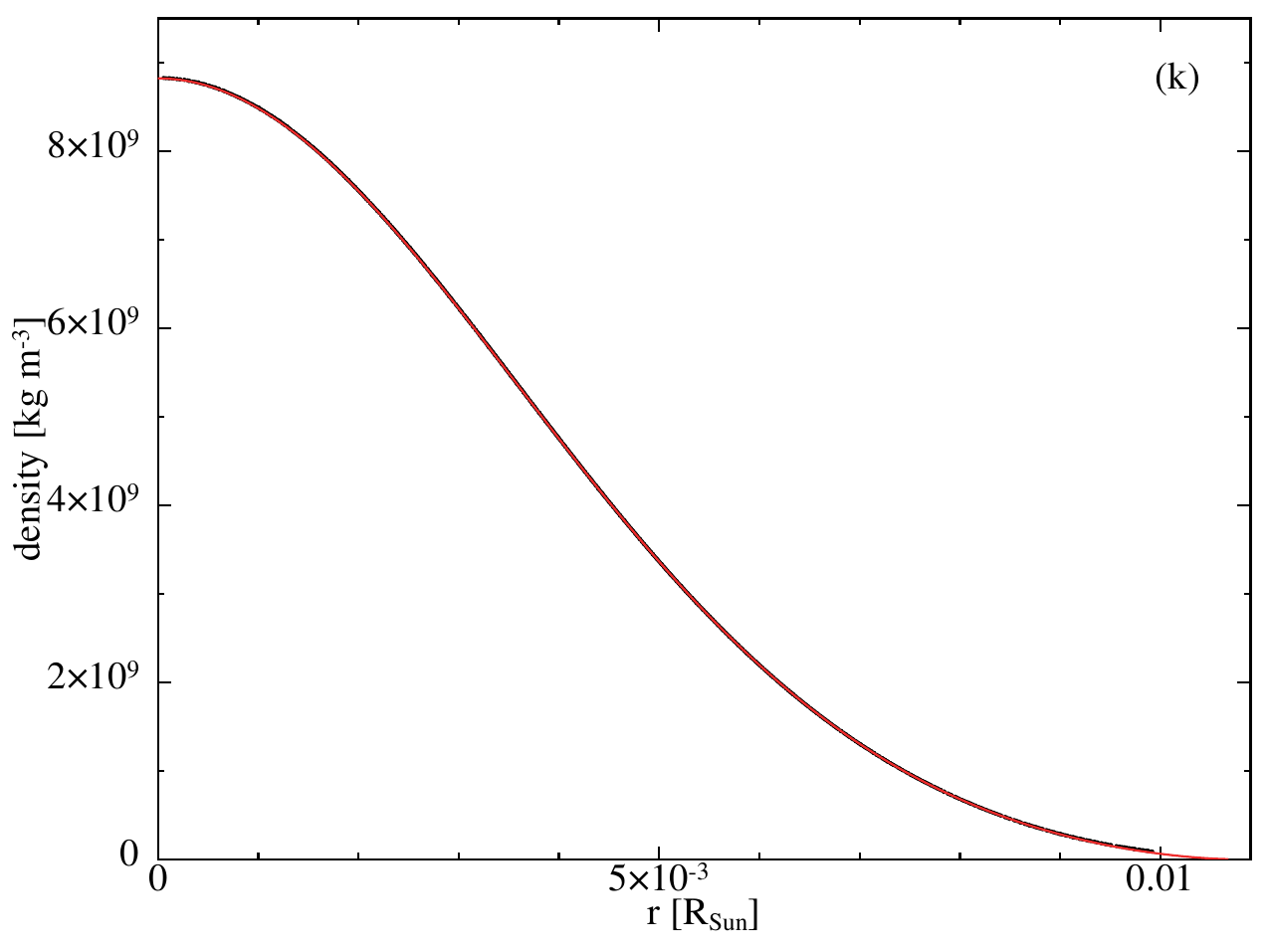}
	\includegraphics[scale=0.24]{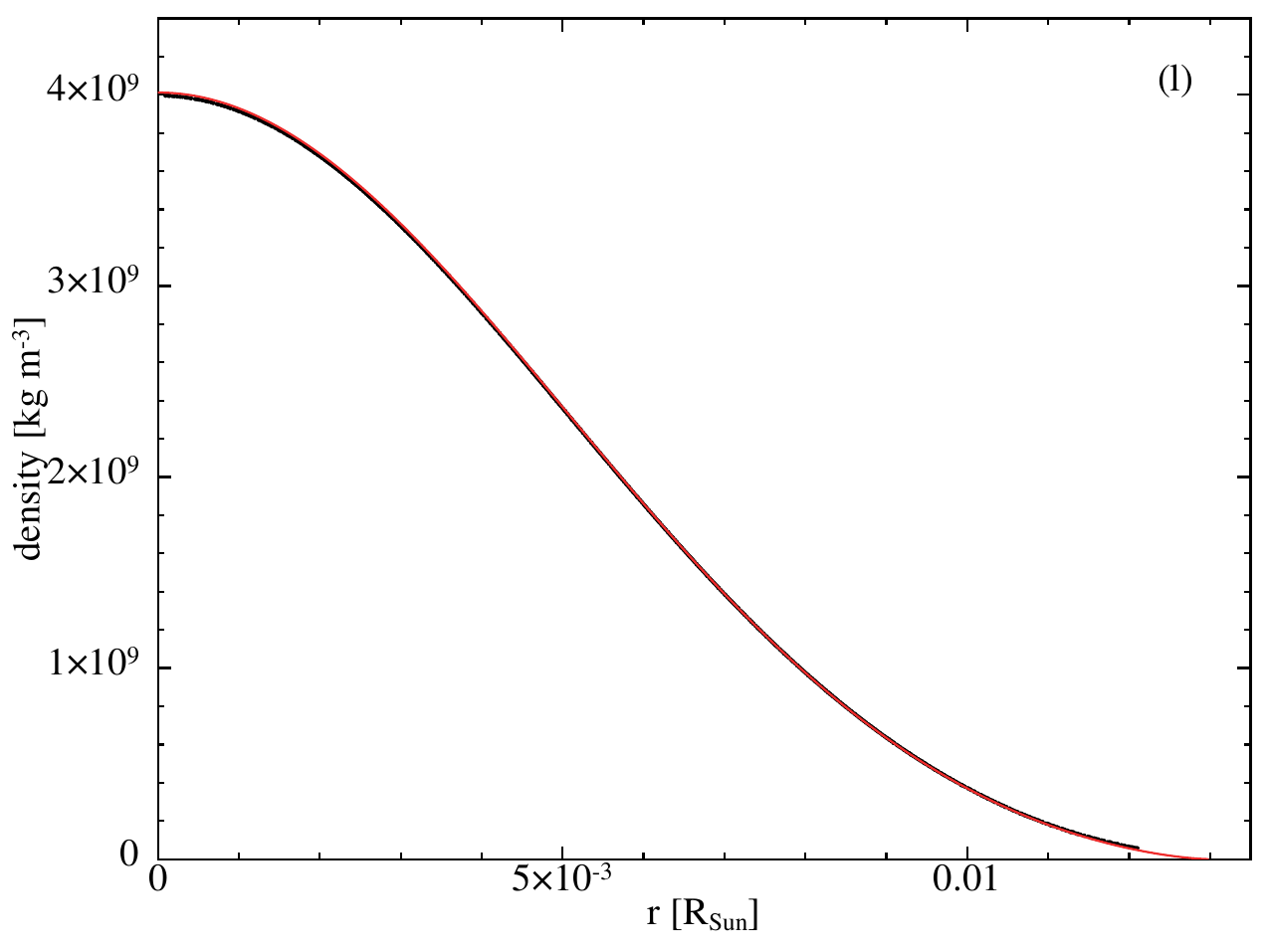}
	
	\caption{The relaxed density profiles are compared with the initial density profiles obtained from the profile $x(r)$ using Equation (\ref{rho_eq}). These figures are generated using SPLASH, with the parameter values for the panels as follows: (a) $M_{\text{wd}} = 0.50 M_{\odot}$, $R_{\text{wd}} = 0.0128 R_{\odot}$, (b) $M_{\text{wd}} = 0.50 M_{\odot}$, $R_{\text{wd}} = 0.0136 R_{\odot}$, (c) $M_{\text{wd}} = 0.50 M_{\odot}$, $R_{\text{wd}} = 0.0141 R_{\odot}$, (d) $M_{\text{wd}} = 0.50 M_{\odot}$, $R_{\text{wd}} = 0.0151 R_{\odot}$, (e) $M_{\text{wd}} = 0.50 M_{\odot}$, $R_{\text{wd}} = 0.0167 R_{\odot}$, (f) $M_{\text{wd}} = 0.75 M_{\odot}$, $R_{\text{wd}} = 0.0098 R_{\odot}$, (g) $M_{\text{wd}} = 0.75 M_{\odot}$, $R_{\text{wd}} = 0.0109 R_{\odot}$, (h) $M_{\text{wd}} = 0.75 M_{\odot}$, $R_{\text{wd}} = 0.0125 R_{\odot}$, (i) $M_{\text{wd}} = 0.75 M_{\odot}$, $R_{\text{wd}} = 0.0144 R_{\odot}$, (j) $M_{\text{wd}} = 1.00 M_{\odot}$, $R_{\text{wd}} = 0.0082 R_{\odot}$, (k) $M_{\text{wd}} = 1.00 M_{\odot}$, $R_{\text{wd}} = 0.0107 R_{\odot}$, (l) $M_{\text{wd}} = 1.00 M_{\odot}$, $R_{\text{wd}} = 0.0130 R_{\odot}$.}
	\label{fig.density}
\end{figure}



\begin{thebibliography}{999}

\bibitem{CliftonRev}
Clifton T., Ferreira P.~G., Padilla A., Skordis C., 2012, \href{https://doi.org/10.1016/j.physrep.2012.01.001}{Physics Reports}, \href{Clifton, Timothy, Ferreira, Pedro G., Padilla, Antonio, & Skordis, Constantinos}{{\bf 513}, 1}.

\bibitem{NOORev}
Nojiri S., Odintsov S.~D., Oikonomou V.~K., 2017, \href{https://doi.org/10.1016/j.physrep.2017.06.001}{Physics Reports}, \href{Nojiri, S., Odintsov, S. D., & Oikonomou, V. K.}{{\bf 692}, 1}. arXiv:1705.11098

\bibitem{LangloisRev}
Langlois D., 2019, \href{https://doi.org/10.1142/S0218271819420069}{International Journal of Modern Physics D}, \href{Langlois, David}{{\bf 28}, 1942006-3287}.

\bibitem{IshakRev} 
Ishak M., 2019, \href{https://doi.org/10.1007/s41114-018-0017-4}{Living Reviews in Relativity}, \href{Ishak, Mustapha}{{\bf 22}, 1}.

\bibitem{KaseRev} 
Kase R., Tsujikawa S., 2019, \href{https://doi.org/10.1142/S0218271819420057}{International Journal of Modern Physics D}, \href{Kase, Ryotaro & Tsujikawa, Shinji}{{\bf 28}, 1942005}.

\bibitem{Penrose} 
Penrose R., 1965, \href{https://doi.org/10.1103/PhysRevLett.14.57}{Physical Review Letters}, \href{Penrose, Roger}{{\bf 14}, 57}.

\bibitem{Hawking} 
Hawking S.~W., 1966, \href{https://doi.org/10.1103/PhysRevLett.17.444}{Physical Review Letters}, \href{Hawking, S. W.}{{\bf 17}, 444}.

\bibitem{HawkingEllis} 
Hawking S.~W., Ellis G.~F.~R., 1973, ‘The Large Scale Structure of Space-Time’, Cambridge University Press.

\bibitem{Wald} 
Wald R.~M., 1984, ‘General Relativity’, The University of Chicago Press.

\bibitem{Senovilla} 
Senovilla J.~M.~M., 1998, \href{https://doi.org/10.1023/A:1018801101244}{General Relativity and Gravitation}, \href{Senovilla, José M. M.}{{\bf 30}, 701}.

\bibitem{SaksteinRev}
Baker T., Barreira A., Desmond H., Ferreira P., Jain B., Koyama K., Li B., Lombriser L., Nicola A., Sakstein J., Schmidt F., 2021, \href{https://doi.org/10.1103/RevModPhys.93.015003}{Reviews of Modern Physics}, \href{Baker, Tessa, Barreira, Alexandre, Desmond, Harry, Ferreira, Pedro, Jain, Bhuvnesh, Koyama, Kazuya, Li, Baojiu, Lombriser, Lucas, Nicola, Andrina, Sakstein, Jeremy, & Schmidt, Fabian}{{\bf 93}, 015003}.

\bibitem{OlmoRev1}
Olmo G.~J., Rubiera-Garcia D., Wojnar A., 2020, \href{https://doi.org/10.1016/j.physrep.2020.07.001}{Physics Reports}, \href{Olmo, Gonzalo J., Rubiera-Garcia, Diego, & Wojnar, Aneta}{{\bf 876}, 1}. arXiv:1912.05202

\bibitem{Born1} 
Born M., 1933, \href{https://doi.org/10.1038/132282a0}{Nature}, \href{Born, M.}{{\bf 132}, 282}.

\bibitem{Born2} 
Born M., 1934, \href{https://doi.org/10.1098/rspa.1934.0010}{Proceedings of the Royal Society of London Series A}, \href{Born, Max}{{\bf 143}, 410}.

\bibitem{BI1} 
Born M., Infeld L., 1934, \href{https://doi.org/10.1098/rspa.1934.0059}{Proceedings of the Royal Society of London Series A}, \href{Born, M. & Infeld, L.}{{\bf 144}, 425}.

\bibitem{BF} 
Banados M., Ferreira P.~G., 2014, \href{https://doi.org/10.1103/PhysRevLett.113.119901}{Physical Review Letters}, \href{Bañados, Máximo & Ferreira, Pedro G.}{{\bf 113}, 119901}.


\bibitem{DG} 
Deser S., Gibbons G.~W., 1998, \href{https://doi.org/10.1088/0264-9381/15/5/001}{Classical and Quantum Gravity}, \href{Deser, S. & Gibbons, G. W.}{{\bf 15}, L35}. arXiv:hep-th/9803049


\bibitem{Vollick1} 
Vollick D.~N., 2004, \href{https://doi.org/10.1103/PhysRevD.69.064030}{Physical Review D}, \href{Vollick, Dan N.}{{\bf 69}, 064030}. arXiv:gr-qc/0309101


\bibitem{Vollick2} 
Vollick D.~N., 2005, \href{https://doi.org/10.1103/PhysRevD.72.084026}{Physical Review D}, \href{Vollick, Dan N.}{{\bf 72}, 084026}. arXiv:gr-qc/0506091


\bibitem{Eddington} 
Eddington A.~S., 1924, ‘The Mathematical Theory of Relativity’, Cambridge University Press.

\bibitem{Schrodinger} 
Schrodinger E., 1985, ‘Space-time Structure’, Cambridge University Press.
 
\bibitem{new4} 
Jimenez J.~B., Delhom A., 2020, \href{https://doi.org/10.1140/epjc/s10052-020-8143-z}{European Physical Journal C}, \href{Jiménez, Jose Beltrán & Delhom, Adrià}{{\bf 80}, 585}. arXiv:2004.11357


\bibitem{Pani1} 
Pani P., Cardoso V., Delsate T., 2011, \href{https://doi.org/10.1103/PhysRevLett.107.031101}{Physical Review Letters}, \href{Pani, Paolo, Cardoso, Vitor, & Delsate, Térence}{{\bf 107}, 031101}. arXiv:1106.3569


\bibitem{Delsate1} 
Delsate T., Steinhoff J., 2012, \href{https://doi.org/10.1103/PhysRevLett.109.021101}{Physical Review Letters}, \href{Delsate, Térence & Steinhoff, Jan}{{\bf 109}, 021101}. arXiv:1201.4989


\bibitem{Casanellas1} 
Casanellas J., Pani P., Lopes I., Cardoso V., 2012, \href{https://doi.org/10.1088/0004-637X/745/1/15}{The Astrophysical Journal}, \href{Casanellas, Jordi, Pani, Paolo, Lopes, Ilídio, & Cardoso, Vitor}{{\bf 745}, 15}. arXiv:1109.0249


\bibitem{Avelino1} 
Avelino P.~P., 2012, \href{https://doi.org/10.1103/PhysRevD.85.104053}{Physical Review D}, \href{Avelino, P. P.}{{\bf 85}, 104053}. arXiv:1201.2544


\bibitem{Avelino2} 
Avelino P.~P., 2012, \href{https://doi.org/10.1088/1475-7516/2012/11/022}{Journal of Cosmology and Astroparticle Physics}, \href{Avelino, P. P.}{{\bf 11}, 022}. arXiv:1207.4730


\bibitem{Banerjee2017} Banerjee S., Shankar S., Singh T.~P., 2017, \href{https://doi.org/10.1088/1475-7516/2017/10/004}{Journal of Cosmology and Astroparticle Physics}, \href{Banerjee, Srimanta, Shankar, Swapnil, & Singh, Tejinder P.}{{\bf 10}, 004}. arXiv:1705.01048

\bibitem{Olmo1} 
Olmo G.~J., Rubiera-Garcia D., Sanchez-Puente A., 2015, \href{https://doi.org/10.1103/PhysRevD.92.044047}{Physical Review D}, \href{Olmo, Gonzalo J., Rubiera-Garcia, D., & Sanchez-Puente, A.}{{\bf 92}, 044047}. arXiv:1508.03272


\bibitem{Pani2012}
Pani P., Delsate T., Cardoso V., 2012, \href{https://doi.org/10.1103/PhysRevD.85.084020}{Physical Review D}, \href{Pani, Paolo, Delsate, Térence, & Cardoso, Vitor}{{\bf 85}, 084020}. arXiv:1201.2814



\bibitem{new1} 
Delhom-Latorre A., Olmo G.~J., Ronco M., 2018, \href{https://doi.org/10.1016/j.physletb.2018.03.002}{Physics Letters B}, \href{Delhom-Latorre, Adrià, Olmo, Gonzalo J., & Ronco, Michele}{{\bf 780}, 294}. arXiv:1709.04249


\bibitem{new5} 
Feng W.-X., Geng C.-Q., Luo, L.-W., 2019, \href{https://doi.org/10.1088/1674-1137/43/8/083107}{Chinese Physics C}, \href{Feng, Wei-Xiang, Geng, Chao-Qiang, & Luo, Ling-Wei}{{\bf 43}, 083107}. arXiv:1810.06753

\bibitem{Rosyadi2019}
Rosyadi A.~S., Sulaksono A., Kassim H.~A., Yusof N., 2019, \href{https://doi.org/10.1140/epjc/s10052-019-7560-3}{European Physical Journal C}, \href{Rosyadi, A. S., Sulaksono, A., Kassim, H. A., & Yusof, N.}{{\bf 79}, 1030}.


\bibitem{new2} 
Delhom A., Miralles V., Penuelas A., 2020, \href{https://doi.org/10.1140/epjc/s10052-020-7880-3}{European Physical Journal C}, \href{Delhom, Adrià, Miralles, Victor, & Peñuelas, Ana}{{\bf 80}, 340}. arXiv:1907.05615

\bibitem{new3} 
Jimenez J.~B., Delhom A., Olmo G.~J., Orazi E., 2021, \href{https://doi.org/10.1016/j.physletb.2021.136479}{Physics Letters B}, \href{Jiménez, Jose Beltrán, Delhom, Adrià, Olmo, Gonzalo J., & Orazi, Emanuele}{{\bf 820}, 136479}.

\bibitem{Banerjee2022} 
Banerjee P., Garain D., Paul S., Shaikh R., Sarkar T., 2022, \href{https://doi.org/10.3847/1538-4357/ac324f}{The Astrophysical Journal}, \href{Banerjee, Pritam, Garain, Debojyoti, Paul, Suvankar, Shaikh, Rajibul, & Sarkar, Tapobrata}{{\bf 924}, 20}. arXiv:2105.09172


\bibitem{OlmoRev} 
Jimenez J.~B., Heisenberg L., Olmo G.~J., Rubiera-Garcia D., 2018, \href{https://doi.org/10.1016/j.physrep.2017.11.001}{Physics Reports}, \href{Beltrán Jiménez, Jose, Heisenberg, Lavinia, Olmo, Gonzalo J., & Rubiera-Garcia, Diego}{{\bf 727}, 1}. arXiv:1704.03351


\bibitem{KobayashiRev}
Kobayashi T., 2019, \href{https://doi.org/10.1088/1361-6633/ab2429}{Reports on Progress in Physics}, \href{Kobayashi, Tsutomu}{{\bf 82}, 086901}. arXiv:1901.07183


\bibitem{Hills1975}
Hills J.~G., 1975, \href{https://doi.org/10.1038/254295a0}{Nature}, \href{Hills, J. G.}{{\bf 254}, 295}.


\bibitem{Maguire2020}
Maguire K., Eracleous M., Jonker P.~G., MacLeod M., Rosswog S., 2020, \href{https://doi.org/10.1007/s11214-020-00661-2}{Space Science Reviews}, \href{Maguire, Kate, Eracleous, Michael, Jonker, Peter G., MacLeod, Morgan, & Rosswog, Stephan}{{\bf 216}, 39}.

\bibitem{Koester1990}
Koester D., Chanmugam G., 1990, \href{https://doi.org/10.1088/0034-4885/53/7/001}{Reports on Progress in Physics}, \href{Koester, D. & Chanmugam, G.}{{\bf 53}, 837}.

\bibitem{Isern2022}
Isern J., Torres S., Rebassa-Mansergas A., 2022, \href{https://doi.org/10.3389/fspas.2022.815517}{Frontiers in Astronomy and Space Sciences}, \href{Isern, J., Torres, S., & Rebassa-Mansergas, A.}{{\bf 9}, 6}.

\bibitem{Fontaine2013}
Fontaine G., Brassard P., Charpinet S., Randall S.~K., Van Grootel V., 2013, \href{https://doi.org/10.1051/epjconf/20134305001}{European Physical Journal Web of Conferences}, \href{Fontaine, G., Brassard, P., Charpinet, S., Randall, S. K., & Van Grootel, V.}{{\bf 43}, 05001}.

	
\bibitem{shapiro} Shapiro S.~L., Teukolsky S.~A., 1983, ‘Black Holes, WDs, and Neutron Stars: The Physics of Compact Objects’, Wiley.

\bibitem{Jain}
Jain R.~K., Kouvaris C., Nielsen N.~G., 2016, \href{https://doi.org/10.1103/PhysRevLett.116.151103}{Physical Review Letters}, \href{Jain, Rajeev Kumar, Kouvaris, Chris, & Nielsen, Niklas Grønlund}{{\bf 116}, 151103}. arXiv:1512.05946

\bibitem{Boshkayev}
Boshkayev K., 2018, \href{https://doi.org/10.1134/S106377291812017X}{Astronomy Reports}, \href{Boshkayev, K.}{{\bf 62}, 847}. arXiv:1807.00332


\bibitem{Holberg2012}
Holberg J.~B., Oswalt T.~D., Barstow M.~A., 2012, \href{https://doi.org/10.1088/0004-6256/143/3/68}{The Astronomical Journal}, \href{Holberg, J. B., Oswalt, T. D., & Barstow, M. A.}{{\bf 143}, 68}. arXiv:1201.3822

\bibitem{Taubenberger}
Taubenberger S., Benetti S., Childress M., Pakmor R., Hachinger S., Mazzali P.~A., Stanishev V., Elias-Rosa N., Agnoletto I., Bufano F., Ergon M., Harutyunyan A., Inserra C., Kankare E., Kromer M., Navasardyan H., Nicolas J., Pastorello A., Prosperi E., Salgado F., Sollerman J., Stritzinger M., Turatto M., Valenti S., Hillebrandt W., 2011, \href{https://doi.org/10.1111/j.1365-2966.2010.18107.x}{Monthly Notices of the Royal Astronomical Society}, \href{Taubenberger, S., Benetti, S., Childress, M., Pakmor, R., Hachinger, S., Mazzali, P. A., Stanishev, V., Elias-Rosa, N., Agnoletto, I., Bufano, F., Ergon, M., Harutyunyan, A., Inserra, C., Kankare, E., Kromer, M., Navasardyan, H., Nicolas, J., Pastorello, A., Prosperi, E., Salgado, F., Sollerman, J., Stritzinger, M., Turatto, M., Valenti, S., & Hillebrandt, W.}{{\bf 412}, 2735}. arXiv:1011.5665



\bibitem{Banerjee2023} Banerjee P., Garain D., Chowdhury S., Singh D., Joshi R., Sarkar T., 2023, \href{https://doi.org/10.1093/mnras/stad1284}{Monthly Notices of the Royal Astronomical Society}, \href{Banerjee, Pritam, Garain, Debojyoti, Chowdhury, Shaswata, Singh, Dhananjay, Joshi, Rohan, & Sarkar, Tapobrata}. arXiv:2212.09122

\bibitem{Herant} Herant M., 1994, Memorie della Societa Astronomica Italiana, \href{https://ui.adsabs.harvard.edu/abs/1994MmSAI..65.1013H}{{\bf 65}, 1013}. 

\bibitem{Coughlin2015}
Coughlin E.~R., Nixon C., 2015, \href{https://doi.org/10.1088/2041-8205/808/1/L11}{The Astrophysical Journal}, \href{Coughlin, Eric R. & Nixon, Chris}{{\bf 808}, L11}. arXiv:1506.08194


\bibitem{Golightly2019} Golightly E.~C.~A., Nixon C.~J., Coughlin E.~R., 2019, \href{https://doi.org/10.3847/2041-8213/ab380d}{The Astrophysical Journal}, \href{Golightly, E. C. A., Nixon, C. J., & Coughlin, E. R.}{{\bf 882}, L26}. arXiv:1907.05895

\bibitem{Miles2020}
Miles P.~R., Coughlin E.~R., Nixon C.~J., 2020, \href{https://doi.org/10.3847/1538-4357/ab9c9f}{The Astrophysical Journal}, \href{Miles, Patrick R., Coughlin, Eric R., & Nixon, C. J.}{{\bf 899}, 36}. arXiv:2006.09375

\bibitem{Cufari2022}
Cufari M., Coughlin E.~R., Nixon C.~J., 2022, \href{https://doi.org/10.3847/1538-4357/ac32be}{The Astrophysical Journal}, \href{Cufari, M., Coughlin, Eric R., & Nixon, C. J.}{{\bf 924}, 34}. arXiv:2110.11374

\bibitem{Mockler2019}
Mockler B., Guillochon J., Ramirez-Ruiz E., 2019, \href{https://doi.org/10.3847/1538-4357/ab010f}{The Astrophysical Journal}, \href{Mockler, Brenna, Guillochon, James, & Ramirez-Ruiz, Enrico}{{\bf 872}, 151}. arXiv:1801.08221


\bibitem{Guillochon2013} Guillochon J., Ramirez-Ruiz E., 2013, \href{https://doi.org/10.1088/0004-637X/767/1/25}{ The Astrophysical Journal}, \href{https://ui.adsabs.harvard.edu/abs/2013ApJ...767...25G}{{\bf 767}, 25}. arXiv:1206.2350 

\bibitem{Manukian2013}
Manukian H., Guillochon J., Ramirez-Ruiz E., O'Leary R. M., 2013, \href{https://doi.org/10.1088/2041-8205/771/2/L28}{The Astrophysical Journal}, \href{Manukian, Haik, Guillochon, James, Ramirez-Ruiz, Enrico, & O'Leary, Ryan M.}{{\bf 771}, L28}. arXiv:1305.4634

\bibitem{Gafton2015}
Gafton E., Tejeda E., Guillochon J., Korobkin O., Rosswog S., 2015, \href{https://doi.org/10.1093/mnras/stv350}{Monthly Notices of the Royal Astronomical Society}, \href{Gafton, E., Tejeda, E., Guillochon, J., Korobkin, O., & Rosswog, S.}{{\bf 449}, 771}. arXiv:1502.02039


\bibitem{PHANTOM} Price D.~J., Wurster J., Tricco T.~S., et al., 2018, \href{https://doi.org/10.1017/pasa.2018.25}{ Publications of the Astronomical Society of Australia}, \href{https://ui.adsabs.harvard.edu/abs/2018PASA...35...31P}{{\bf 35}, e031}. arXiv:1702.03930 

\bibitem{Rees1988}
Rees M.~J., 1988, \href{https://doi.org/10.1038/333523a0}{Nature}, \href{Rees, Martin J.}{{\bf 333}, 523}.


\bibitem{Coughlin2019} Coughlin E.~R., Nixon C.~J., 2019, \href{https://doi.org/10.3847/2041-8213/ab412d}{The Astrophysical Journal}, \href{Coughlin, Eric R. & Nixon, C. J.}{{\bf 883}, L17}. arXiv:1907.03034  

\bibitem{Golightly2} 
Golightly E.~C.~A., Coughlin E.~R., Nixon C.~J., 2019, \href{https://doi.org/10.3847/1538-4357/aafd2f}{The Astrophysical Journal}, \href{Golightly, Elen C. A., Coughlin, Eric R., & Nixon, C. J.}{{\bf 872}, 163}. arXiv:1901.03717

\end{thebibliography}
\end{document}